\documentclass[12pt,a4paper,twoside]{article}

\usepackage{a4}
\usepackage{amsmath}
\usepackage{amssymb}
\usepackage{amsthm}
\usepackage{array}
\IfFileExists{bbm.sty}{\usepackage{bbm}}{\typeout{ * ^^J
   * Please install the bbm bundle! I try to replace the bbm fonts by the ^^J
   * AMS blackboard font and the cm bold font ^^J *}
   \newcommand{\mathbbm}[1]{\mathbb{##1}}}
\IfFileExists{mathrsfs.sty}{\usepackage{mathrsfs}}{\typeout{ * ^^J
   * Please install the mathrsfs bundle! I try to replace the rsfs font by ^^J
   * the cm script font ^^J *}
   \newcommand{\mathscr}[1]{\mathcal{##1}}}
\IfFileExists{cite.sty}{\usepackage{cite}}{}

\makeatletter
\def\preprint#1{\def\@preprint{#1}} \def\@preprint{}
\def\email#1{\def\@email{#1}} \def\@email{}
\def\address#1{\def\@address{#1}} \def\@address{} 
\def\@maketitle{%
  \newpage
  \null
  {}\hfill \raisebox{3cm}{\if \@preprint \empty \else \@preprint \fi}
  \vspace{-2cm}
  \begin{center}%
  \let \footnote \thanks
    {\LARGE\sffamily \@title \par}%
    \vskip 1.5em%
    {\large\scshape
      \lineskip .5em%
      \begin{tabular}[t]{c}%
        \@author
      \end{tabular}\par}%
    \vskip 1em%
    {\if \@address\empty \else \large\slshape
      \lineskip .5em%
      \begin{tabular}[t]{c}%
	\@address
      \end{tabular}\par \fi}%
     \vskip 0.5em%
    {\if \@email\empty \else \large e-mail:\space %
				   \texttt{\ignorespaces \@email} \fi}
     \vskip 1.5em%
    {\large \@date}%
  \end{center}%
  \par
  \vskip 2em}
\makeatother

\makeatletter
\def\eqnarr@left{\@centering}
\let\eqnarr@opts\relax
\def\equationarray{%
 \col@sep\arraycolsep
 \def\d@llarbegin{$\displaystyle}%
 \def\d@llarend{$}%
 \stepcounter{equation}%
 \let\@currentlabel=\theequation
 \set@eqnsw \global\@eqcnt\z@ \global\@eqargcnt\z@
 \let\@classz\@eqnclassz
 \def\multicolumn##1##2##3{\@eqnmulticolumn{##1}{##2}{##3}%
                           \global\advance\@eqcnt##1
                           \global\advance\@eqcnt\m@ne}%
 \def\@halignto{to\displaywidth}%
 \@ifnextchar[{\@equationarray}{\@equationarray[.]}}
\let\@eqnmulticolumn=\multicolumn
\def\yesnumber{\global\@eqnswtrue}
\let\set@eqnsw=\yesnumber
\def\@amper{&}
\newcount\@eqargcnt  
\def\@equationarray[#1]#2{%
     \eqnarr@opts
     \@tempdima \ht \strutbox
     \advance \@tempdima by\extrarowheight
     \setbox\@arstrutbox=\hbox{\vrule
               \@height\arraystretch \@tempdima
               \@depth\arraystretch \dp \strutbox
               \@width\z@}%
     \gdef\advance@eqargcnt{\global\advance\@eqargcnt\@ne}%
     \begingroup
     \@mkpream{#2}%
     \xdef\@preamble{%
      \if #1l\tabskip\z@ \else\if #1r\tabskip\@centering
                         \else\if #1c\tabskip\@centering
                         \else\tabskip\eqnarr@left \fi\fi\fi
      \halign \@halignto
      \bgroup \tabskip\z@ \@arstrut \@preamble
      \if #1l\tabskip\@centering \else\if #1r\tabskip\z@
                                 \else\tabskip\@centering \fi\fi
      \@amper\llap{\@sharp}\tabskip\z@\cr}%
     \endgroup
     \gdef\advance@eqargcnt{}%
     \bgroup
     \let\@sharp## \let\protect\relax
     \m@th   \let\\=\@equationcr
     \let\par\@empty
     $$                            
     \lineskip \z@
     \baselineskip \z@
     \@preamble}
\def\@eqnclassz{\@classx
   \@tempcnta \count@
   \advance@eqargcnt
   \prepnext@tok
   \@addtopreamble{%
      \global\advance\@eqcnt\@ne
      \ifcase \@chnum
      \hfil \d@llarbegin \insert@column \d@llarend\hfil \or
      \d@llarbegin \insert@column \d@llarend \hfil \or
      \hfil\kern\z@ \d@llarbegin \insert@column \d@llarend \or
      $\vcenter
      \@startpbox{\@nextchar}\insert@column \@endpbox $\or
      \vtop \@startpbox{\@nextchar}\insert@column \@endpbox \or
      \vbox \@startpbox{\@nextchar}\insert@column \@endpbox
      \fi}\prepnext@tok}
\def\endequationarray{\@zequationcr
   \egroup
   \global\advance\c@equation\m@ne $$  
   \egroup\global\@ignoretrue
   \gdef\@preamble{}}
\def\@equationcr{${\ifnum0=`}\fi\@ifstar{\global\@eqpen\@M
    \@xequationcr}{\global\@eqpen\interdisplaylinepenalty
                   \@xequationcr}}
\def\@xequationcr{%
    \@ifnextchar[{\@argequationcr}{\ifnum0=`{\fi}${}%
    \@zequationcr}}
\def\@argequationcr[#1]{\ifnum0=`{\fi}${}\ifdim #1>\z@
   \@xargequationcr{#1}\else
   \@yargequationcr{#1}\fi}
\def\@xargequationcr#1{\unskip
   \@tempdima #1\advance\@tempdima \dp \@arstrutbox
   \vrule \@depth\@tempdima \@width\z@
   \@zequationcr\noalign{\penalty\@eqpen}}
\def\@yargequationcr#1{%
   \@zequationcr\noalign{\penalty\@eqpen\vskip #1}}
\def\@zequationcr{\@whilenum\@eqcnt <\@eqargcnt
   \do{\@amper\omit\global\advance\@eqcnt\@ne}%
   \@amper
   \if@eqnsw\@eqnnum\stepcounter{equation}\fi
   \set@eqnsw\global\@eqcnt\z@\cr}
\@namedef{equationarray*}{%
   \let\set@eqnsw=\nonumber \equationarray}
\@namedef{endequationarray*}{\endequationarray}
\makeatother

\DeclareMathOperator{\id}{id}
\DeclareMathOperator{\End}{End}

\DeclareMathOperator{\tr}{tr}
\DeclareMathOperator{\Tr}{Tr}

\DeclareMathOperator{\grad}{grad}

\def\ot{\otimes}
\def\op{\oplus}
\def\t{\tilde}
\def\h{\hat}
\def\iu{\mathrm{i}}
\def\vg{\mathrm{v}_{\!g}}
\def\ctr{\;\vrule width0.35em height0.10ex depth0ex 
	   \vrule width0.05em height1.50ex depth0ex \;}
 
\def\bfd{\mathbf{d}}
\def\sfD{\mathsf{D}}
\def\sfd{\mathsf{d}}
\def\bsj{\boldsymbol{\psi}}
\def\bsw{\boldsymbol{\omega}}
\def\ga{\boldsymbol{\gamma}}
\def\vrs{\varsigma}

\def\vpi{\varpi}
\def\vrr{\varrho}
\def\rmf{\mathrm{f}}
\def\rmtf{\tilde{\mathrm{f}}}
\def\one{\mathbbm{1}}
\def\CX{{C^\infty (X)}}
\def\B{\mathscr{B}}
\def\C{\mathbbm{C}}
\def\N{\mathbbm{N}}
\def\M{\mathcal{M}}
\def\R{\mathbbm{R}}
\def\Z{\mathbbm{Z}}

\def\SU#1{\mathrm{SU(#1)}}
\def\U#1{\mathrm{U(#1)}}

\def\u#1{\mathrm{u(#1)}}

\def\yn{\yesnumber}
\def\npb{\nopagebreak}
\newcommand\eq[2][]{\begin{equation} \label{#1} #2 \end{equation}}
\newcommand\eqa[2]{\begin{equationarray*}{#1} #2 \end{equationarray*}}
\newcommand\eqas[3][]{\begin{equation}\label{#1} \begin{array}{#2} #3 
		      \end{array} \end{equation}}
\newcommand\al[1]{\begin{align} #1 \end{align}}
\newcommand\aln[1]{\begin{align*} #1 \end{align*}}

\newcommand\seq[1]{\begin{subequations} #1 \end{subequations}}

\newcommand{\cmcal}[1]{\mathcal{#1}}
\newcommand{\W}[2][g]{\ensuremath{\Omega^{#2}{\mathfrak #1}}}
\newcommand{\tW}[2][g]{\ensuremath{\tilde{\Omega}^{#2}{\mathfrak #1}}}
\renewcommand{\P}[2][g]{\ensuremath{\pi(\Omega^{#2}{\mathfrak #1})}}
\newcommand{\p}[2][a]{\ensuremath{\hat{\pi}(\Omega^{#2}{\mathfrak #1})}}
\newcommand{\D}[2][g]{\ensuremath{\Omega_D^{#2} \mathfrak{#1}}}
\newcommand{\J}[2][g]{\ensuremath{\pi(\mathcal{J}^{#2}\!\mathfrak{#1})}}
\newcommand{\cJ}[2][g]{\ensuremath{\mathbbm{J}^{#2}\!\mathfrak{#1}}}
\newcommand{\cj}[2][a]{\ensuremath{\mathbbm{j}^{#2}\!\mathfrak{#1}}}
\newcommand{\bbr}[2][a]{\ensuremath{\mathbbm{r}^{#2}\!\mathfrak{#1}}}
\newcommand{\jj}[2][a]{\ensuremath{\hat{\pi}(\mathcal{J}^{#2} \mathfrak{#1})}}
\newcommand{\pf}[1][a]{\ensuremath{\hat{\pi}({\mathfrak #1})}}
\newcommand{\K}[2]{\ensuremath{\hat{\pi}(T^{#1}_{#2} \mathfrak{a})}}

\newcommand{\tJ}[3][a]{\ensuremath{\tilde{\mathcal{J}\,}\!^{#2}_{#3}%
				  {\mathfrak #1}}}
\newcommand{\tK}[3][a]{\ensuremath{\tilde{K}^{#2}_{#3}{\mathfrak #1}}}
\newcommand{\hH}[2][g]{\ensuremath{\h{\cmcal{H}}^{#2}{\mathfrak #1}}}
\newcommand{\tH}[2][g]{\ensuremath{\t{\cmcal{H}}^{#2}{\mathfrak #1}}}
\renewcommand{\H}[2][g]{\ensuremath{\cmcal{H}^{#2}{\mathfrak #1}}}
\newcommand{\tcc}[2][g]{\tilde{\mathbbm{c}}^{#2}\!\mathfrak{#1}}
\newcommand{\cc}[2][a] {\mathbbm{c}^{#2}\!\mathfrak{#1}}

\def\mc{\multicolumn}
\newcommand{\f}[1][g]{\ensuremath{\mathfrak{#1}}}
\newcommand{\mat}[2][\C]{{\mathrm{M}}_{#2}{#1}}
\newcommand{\rf}[1][]{\textup{\eqref{#1}}}
\newcommand{\g}[1][5]{\gamma^{#1}}
\newcommand{\Ad}[1]{\mathrm{Ad}_{#1}\,}
\newcommand{\miss}[1]{\stackrel{\stackrel{#1}{\vee}}{\dots}}
\newcommand{\hsg}[1][\mathfrak{g}]{\h{\sigma}_{#1}} 

\def\hs{\hspace}
\def\vs{\vspace}
\def\ds{\displaystyle} 
\def\ts{\textstyle}
\def\th{\tfrac{1}{2}}

\def\tsum{\textstyle \sum}
\def\dsum{\displaystyle \sum}

\newtheorem{dfn}{Definition}
\newtheorem{prp}[dfn]{Proposition}
\newtheorem{lem}[dfn]{Lemma}
\newtheorem{thm}[dfn]{Theorem}
\newtheorem{cor}[dfn]{Corollary}
\makeatletter
\renewenvironment{proof}[1][\proofname]{\par
  \normalfont\topsep6\p@\@plus6\p@ \trivlist
  \item[\hskip\labelsep\bfseries #1\@addpunct{:}]\ignorespaces}{%
  \qed\endtrivlist}
\makeatother
\newcommand{\Proof}[2][\proofname]{\begin{proof}[#1] #2 \end{proof}}

\renewcommand{\arraystretch}{1.2}
\numberwithin{equation}{section}

\begin{document}

\pagestyle{plain}

\title {   The Mathematical Footing of Non--associative Geometry}
\author{   Raimar Wulkenhaar} 
\address{  Institut f\"ur Theoretische Physik \\
	   Universit\"at Leipzig \\ 
	   Augustusplatz 10/11, D--04109 Leipzig, Germany} 
\email{    wulkhaar@tph100.physik.uni-leipzig.de} 
\date{	   October 29, 1996}	   
\preprint{ \begin{tabular}{l}hep-th/9607094 \\revised version\end{tabular}}

\maketitle 

\vfill
\vfill

\begin{abstract}
\noindent
Starting with a Hilbert space endowed with a representation of a unitary Lie 
algebra and an action of a generalized Dirac operator, we develop a 
mathematical concept towards gauge field theories. This concept shares common 
features with the non--commutative geometry \`{a} la Connes\,/\,Lott, differs 
from that, however, by the implementation of unitary Lie algebras instead of 
associative $*$--algebras. The general scheme is presented in detail and is 
applied to functions $\otimes$ matrices. 
\end{abstract}

\vfill
\clearpage 

\section{Introduction}

We present a framework towards a construction (of the classical action) of 
gauge field theories out of the following input data:
\vs{-\topsep}
\begin{enumerate}
\item
The (Lie) group of local gauge transformations $\mathscr{G}\,.$  \label{sybr1}
\vs{-\itemsep} \vs{-\parsep}
\item 
Chiral fermions $\bsj$ transforming under a representation $\t{\pi}$ 
of $\mathscr{G}\,.$ 
\vs{-\itemsep} \vs{-\parsep}
\item
The fermionic mass matrix $\widetilde{\M}\,,$ i.e.\ fermion masses plus 
generalized Kobayashi--Maskawa matrices. 
\label{sybr3}
\vs{-\itemsep} \vs{-\parsep}
\item
Possibly the symmetry breaking pattern of $\mathscr{G}\,.$  \label{sybr4}
\end{enumerate}
\vs{-\topsep}
At first sight, this setting seems to be adapted to non--commutative 
geometry \cite{ac}. However, it was proved in \cite{lmms} that only the 
standard model can be constructed within the strictest (and most elegant) 
prescription of non--commutative geometry (NCG)~-- out of a K--cycle 
\cite{ac,cl} with real structure \cite{acr}. For details of this construction 
see \cite{iks, mgv}. Since the standard model is not finally confirmed by 
experiments yet, it must be admitted to consider different physical models such 
as Grand Unified Theories. It is clear from \cite{lmms} that the discussion of 
such models within NCG requires additional structures or different methods. The 
perhaps most successful NCG--approach towards Grand Unification was proposed in 
\cite{cff1, cff2, cf}, where the K--cycle plays an auxiliary r\^ole. 

The author of this paper has sketched in \cite{rw1} a concept towards gauge 
field theories based upon unitary Lie algebras instead of unital associative 
$*$--algebras. Our concept requires the same amount of structures as the 
strictest NCG--setting and is physically motivated. 
Starting from the above physical data 1) to 4) one obtains a K--cycle by 
enlarging the gauge group $\mathscr{G}$ to a unital associative $*$--algebra 
$\mathcal{A}\,,$ provided that it is possible to extend the representation 
$\t{\pi}$ to a representation of $\mathcal{A}\,.$ We shall go the opposite way: 
We restrict the gauge group to its infinitesimal elements, giving the 
Lie algebra of $\mathscr{G}\,.$ In our case there are (apart from 
$\mathrm{U(1)}$--groups) no obstructions for the representation, and~-- in 
principle~-- any physical model based upon 1) to 4) can be constructed.

In this paper we present the mathematical footing of that line, 
which carries the working title ``non--associative geometry''. The reason is 
that Lie algebras are anti--commutative non--associative algebras. We shall 
develop techniques adapted to this case that differ from those of NCG.
Our method can not be applied to general non--associative algebras. Thus, the 
title is somewhat misleading, but  ``anti--commutative non--associative 
geometry'' is too long. 

The paper is organized as follows: Section~\ref{mato} contains the general 
construction, without any reference to a physical model. We start in 
Section~\ref{lcy} with basic definitions concerning L--cycles, the fundamental 
object in non--associative geometry. In Section~\ref{tudla} we construct the 
universal graded differential Lie algebra $\W{*}$ and derive properties of its 
elements. Using the data specified in the L--cycle we define in 
Section~\ref{repr} a Lie algebra representation $\pi$ of $\W{*}$ in $\B(h)\,.$ 
Factorization of $\P{*}$ with respect to the differential ideal $\J{*}$ yields 
the graded differential Lie algebra $\D{*}\,.$ In Section~\ref{tad} we 
introduce the important map $\sigma\,,$ which enables us to give a convenient 
form to the ideal $\J{*}\,.$ Using the language of graded Lie homomorphisms 
introduced in Section~\ref{glh} we define in Section~\ref{secc} the fundamental 
objects of gauge field theories: connections, curvatures, gauge 
transformations, bosonic and fermionic actions.

In Section~\ref{ftm} we apply the general scheme to L--cycles over functions 
$\ot$ matrices. This class of L--cycles, which has a direct relation to 
physical models, is defined in Section~\ref{nota}. For the space--time part it 
is convenient to redefine the exterior differential algebra $\Lambda^*\,,$ see 
Section~\ref{tech}. This enables us decompose in Section~\ref{da} the graded 
Lie algebra $\P{*}$ and in Section~\ref{mthm} the ideal $\J{*}$ into 
space--time part and matrix part. The decomposition of the formulae for the 
differential and the commutator is given in Section~\ref{cdcr}. Finally, we 
consider in Section~\ref{lona} local connections. 

\section{L--cycles and Graded Differential Lie Algebras} 
\label{mato}

\subsection{The L--cycle} 
\label{lcy}

The fundamental geometric object in non--associative geometry is an
L--cycle, which differs from a K--cycle \cite{ac, cl} used in 
non--commutative geometry by the implementation of unitary Lie algebras 
instead of unital associative $*$--algebras: 

\begin{dfn} 
An L--cycle $(\f[g],h,D,\pi,\Gamma)$ over a unitary Lie algebra $\f[g]$ is 
given by 
\\
$\mathrm{i)}$ \hfill \parbox[t]{0.95\textwidth}{an involutive representation 
$\pi$ of $\f$ in the Lie algebra $\B(h)$ of bounded operators on a Hilbert 
space $h\,,$ i.e. $(\pi(a))^*=\pi(a^*) \equiv -\pi(a) \,,$ for any 
$a \in \f\,,$ } 
\\
$\mathrm{ii)}$ \hfill \parbox[t]{0.95\textwidth}{a (possibly unbounded) 
selfadjoint operator $D$ on $h$ such that $(\id_h+D^2)^{-1}$ is compact and 
for all $a \in \f$ there is $[D,\pi(a)] \in \B(h)\,,$ where $\id_h$ denotes 
the identity on $h\,.$} 
\\
$\mathrm{iii)}$ \hfill \parbox[t]{0.95\textwidth}{a selfadjoint operator 
$\Gamma$ on $h\,,$ fulfilling $\Gamma^2=\id_h\,,$ $\Gamma D + D \Gamma=0$ and 
$\Gamma \pi(a)-\pi(a)\Gamma=0\,,$ for all $a \in \f\,.$ } 
\label{lc} 
\end{dfn} 
\noindent
Any Lie algebra $\f[g]$ can be embedded into its universal 
enveloping algebra $\f[U](\f[g])\,,$ and the representation $\pi: \f[g] \to
\B(h)$ extends to a representation $\pi: \f[U](\f[g]) \to \B(h)$ 
(Poincar\'e--Birkhoff--Witt theorem, see \cite{h}). In this sense, any 
L--cycle can be embedded into its ``enveloping K--cycle''. However, the gauge 
field theory obtained by the Connes--Lott prescription \cite{cl, ac} from this 
enveloping K--cycle differs from the gauge field theory we are going to 
develop for the L--cycle. Our construction follows the ideas of Connes and 
Lott, but the methods and results are different. 

Although we do not need it, let us translate properties of a K--cycle into 
definitions for the L--cycle. We use the definition of the distance 
on a K--cycle \cite{ac, cl} to define the distance between linear functionals 
$\mathrm{x}_1,\mathrm{x}_2:\f[g] \to \C$ of the Lie algebra: 
\begin{dfn} 
Let $\mathrm{X}$ be the space of linear functionals of $\f\,.$ The distance 
$\mathrm{dist}(\mathrm{x}_1,\mathrm{x}_2)$ between $\mathrm{x}_1,\mathrm{x}_2 
\in \mathrm{X}$ is given by 
\eqa{rcl}{ 
\mathrm{dist}(\mathrm{x}_1,\mathrm{x}_2):=\sup_{a \in \f} 
\{~|\mathrm{x}_1(a) -\mathrm{x}_2(a)|~:~~ \|\,[D, \pi(a)] \,\| \leq 1 ~\}~. 
	 }
\end{dfn}
\noindent
This definition makes $(\mathrm{X},\mathrm{dist})$ to a metric space, and there 
is no need for $\pi$ being an algebra homomorphism. 

Next, we can take the definition of integration on a K--cycle \cite{ac,cl}
to define the notion of integration on an L--cycle:
\begin{dfn}
Let $\mathrm{d} \in \lbrack 1,\infty)$ be a real number. An L--cycle 
$(\f,h,D,\pi,\Gamma)$ is called $\mathrm{d}^+$--summable if the eigenvalues 
$E_n$ of $D$~-- arranged in increasing order~-- satisfy 
\[
\tsum_{n=1}^N E^{-1}_n = O(\, \tsum_{n=1}^N n^{-1/\mathrm{d}} \,)~. 
\]
We define the integration of $a \in \f$ over $\mathrm{X}$ by 
\[
\int_\mathrm{X} a \,d \mu := \mathit{const.}(\mathrm{d}) \, 
\Tr_{\omega} (\pi(a)|D|^{-\mathrm{d}})~, 
\]
where $\Tr_{\omega}$ is the Dixmier trace, $d \mu$ is the ``volume measure'' on 
$\mathrm{X}$ and $\mathit{const.}(\mathrm{d})$ refers to a constant depending 
on $\mathrm{d}\,.$ \label{d+s} 
\end{dfn}

\subsection{The Universal Graded Differential Lie Algebra $\W{*}$}
\label{tudla} 

To construct differential algebras over a K--cycle $(\mathcal{A},h,D)$ one 
starts from the universal differential algebra $\Omega^*\mathcal{A}$ over 
$\mathcal{A}$ and factorizes this differential algebra with respect to a 
differential ideal determined by the representation $\pi$ of 
$\Omega^* \mathcal{A}$ in $\B(h)\,.$ In analogy to this procedure we first 
define a universal differential Lie algebra $\Omega^*\f$ over the Lie algebra 
$\f$ of the L--cycle. Then we define a representation $\pi$ of $\Omega^*\f$ in 
$\B(h)\,.$ Finally, we perform the factorization with respect to the 
differential ideal. 

\subsubsection{The Tensor Algebra $T(\f[g])$}

Let $\f$ be a Lie algebra over $\R$ with involution given by $a^*=-a\,,$ for 
$a \in \f\,.$ The construction of the universal graded differential Lie 
algebra $\W{*}$ over the Lie algebra $\f$ goes as follows: First, let $d\f$ be 
another copy of $\f\,.$ Let $V(\f)$ be the free vector space generated by 
$\f$ and let $V(d\f)$ be the free vector space generated by $d\f\,,$ 
\eqas{rclrcl}{
V(\f) &:=& {\ds \bigoplus_{a \in \f} V_a} ~,~~ & V_a &=&\R~\forall a \in 
\f[g]~,~~ \\
V(d\f) &:=& {\ds \bigoplus_{da \in d\f} V_{da}~,} \qquad{} &
V_{da} &=& \R~ \forall da \in d\f[g] ~.~~
	    }
For a vector space $\f[X]$ we denote by $\delta_x$ the function on $\f[X]\,,$ 
which takes the value $1$ at the point $x \in \f[X]$ and the value $0$ at all 
points $y \neq x\,.$ Then,
\eqas{rcll}{
V(\f) &=& \{~ \tsum_{\alpha} \lambda_{\alpha} \delta_{a_{\alpha}}~,~~ & 
a_{\alpha} \in \f~,~~\lambda_{\alpha} \in \R~\} ~,~~ \\
V(d\f) &=& \{~ \tsum_{\alpha} \lambda_{\alpha} \delta_{da_{\alpha}}~,~~ & 
a_{\alpha} \in \f~,~~\lambda_{\alpha} \in \R~\}~,
	}
where the sums are finite. Let $T(\f)$ be the tensor algebra of $V(\f) \oplus 
V(d\f)\,,$ which carries a natural $\N$--grading structure. We define 
$\deg(v)=0$ for $v \in V(\f)$ and $\deg(v)=1$ for $v \in V(d\f)\,.$ For 
tensor products $v_1 \ot v_2 \ot \dots \ot v_n \in T(\f)\,,$ where each 
$v_i\,,\ i=1,\dots,n\,,$ belongs either to $V(\f)$ or to $V(d\f)\,,$ we define 
\eq{
\deg(v_1 \ot v_2 \ot \dots \ot v_n):= \tsum_{i=1}^n \deg(v_i)~.~~ 
   }
Now we have 
\eq[tbf]{
T(\f)= \bigoplus_{n \in \N_0} T^n(\f)~,~~ T^n(\f):= \{~ t \in T(\f)~,~~ 
\deg(t)=n~\}~.~~  
	}
In particular, we have $T^k(\f) \ot T^l(\f) \subset T^{k+l}(\f)\,.$ 

Next, we regard $T(\f)$ as a graded Lie algebra with graded commutator given by 
\eq[comt]{
[t^k,\t{t}^l]:=t^k \ot \t{t}^l-(-1)^{kl} \, \t{t}^l \ot t^k~,~~ 
t^k \in T^k(\f)~,~~ \t{t}^l \in T^l(\f)~.~~   
	 }
Obviously, one has 
\eqa{rcl}{
1) &\hs*{2em} & [t^k,\t{t}^l] =-(-1)^{kl} [\t{t}^l, t^k] ~,~~ \npb \\ 
2) && [t^k, \lambda \t{t}^l +\t{\lambda} \Tilde{\Tilde{t}}{}^l] 
= \lambda [t^k, \t{t}^l] + \t{\lambda} [t^k, \Tilde{\Tilde{t}}{}^l]~,  
\yn \npb \\ 
3) && (-1)^{km} [t^k,[\t{t}^l,	\Tilde{\Tilde{t}}{}^m]] + (-1)^{lk} [\t{t}^l, 
[\Tilde{\Tilde{t}}{}^m, t^k]] + (-1)^{ml} [\Tilde{\Tilde{t}}{}^m , 
[t^k,\t{t}^l]]=0 ~, 
\hs{3em} 
	}
for $t^k \in T^k(\f)\,,~\t{t}^l, \Tilde{\Tilde{t}}{}^l \in T^l(\f)\,,~ 
\Tilde{\Tilde{t}}{}^m \in T^m(\f)$ and $\lambda, \t{\lambda} \in \R\,.$ 

\subsubsection{Definition of and Structures on $\W{*}$}

Let $\tW{*}=\bigoplus_{n \in \N_0} \tW{n}$ be the $\N$--graded Lie subalgebra 
of $T(\f)$ given by the set of all repeated commutators (in the sense of 
\rf[comt]) of elements of $V(\f)$ and $V(d\f)\,.$ Let $I'(\f)$ be the vector 
subspace of $\tW{*}$ of elements of the following type: 
\eqas[ipri]{ll}{
\lambda \delta_a-\delta_{\lambda a}~, & 
\lambda \delta_{da}-\delta_{d(\lambda a)}~, \\ 
\delta_{a} +\delta_{\t{a}}-\delta_{a+\t{a}}~, &
\delta_{da} +\delta_{d\t{a}}-\delta_{d(a+\t{a})}~, \\
{}[\delta_{a} ,\delta_{\t{a}}]-\delta_{[a,\t{a}]}~, \\	 
{}[\delta_{da} ,\delta_{\t{a}}]+ [\delta_{a} ,\delta_{d\t{a}}] 
- \delta_{d[a,\t{a}]}~,
      }
for $a,\t{a} \in \f$ and $\lambda \in \R\,.$ Obviously, 
\eqa{rcl}{
I(\f):=I'(\f) &+& [V(\f) \op V(d\f),I'(\f)]  \npb \\
&+& [V(\f) \op V(d\f), [V(\f) \op V(d\f), I'(\f)]] + \dots \yn \label{iip}
	}
is an $\N$--graded ideal of $\tW{*}\,,$ $I(\f)=\bigoplus_{n \in \N_0} 
I^n(\f)\,.$ Then, 
\al{
\W{*} &:=\bigoplus_{n \in \N_0} \W{n}~, & \W{n} &:=\tW{n}\,/\, I^n(\f)~,
   }
is an $\N$--graded Lie algebra, with commutator given by 
\eq{
[\vpi+I(\f),\t{\vpi}+I(\f)]:=[\vpi,\t{\vpi}]+I(\f)~, \quad
\vpi,\t{\vpi} \in \tW{*}~.
   }

On $T(\f)$ we define recursively a graded differential as an $\R$--linear 
map $d: T^n(\f) \to T^{n+1}(\f)$ by 
\eqas[datt]{rclrcl}{ 
d(\lambda \delta_a) &:=& \lambda \delta_{da}~, & 
d(\lambda \delta_{da}) &:=& 0~,~~ \\ 
d(\lambda \delta_a \ot t) &:=& \lambda\delta_{da} \ot t 
+ \lambda\delta_{a} \ot d t ~, \qquad{} &
d(\lambda\delta_{da} \ot t) &:=& - \lambda\delta_{da} \ot dt~,~~
	      }
for $a \in \f\,,$ $t \in T(\f)$ and $\lambda \in \R\,.$ From this definition we 
get 
\eqas{rclrcl}{
d^2(\lambda\delta_a) &=& d(\lambda\delta_{da})=0~, \qquad{} & 
d^2(\lambda\delta_{da}) &=& 0~,~~ \\
d^2(\lambda\delta_a \ot t) &=& d(\lambda\delta_{da} \ot t) 
+ d(\lambda\delta_{a} \ot d t) \\ 
   &=& \mc{4}{l}{ - \lambda\delta_{da} \ot dt + \lambda\delta_{da} \ot d t 
   + \lambda\delta_{a} \ot d^2 t = \lambda \delta_{a} \ot d^2 t ~,~~ } \\ 
d^2(\lambda\delta_{da} \ot t) &=& \lambda\delta_{da} \ot d^2t~,
      }
therefore, by induction, $d^2 \equiv 0$ on $T(\f)\,.$ In order to show that $d$ 
is a graded differential we use the following equivalent characterization of 
\rf[datt]: 
\eq[dvv]{
d( v_1 \ot \dots \ot v_n)=\sum_{i=1}^n (-1)^{\sum_{j=1}^{i-1} \deg(v_j)} 
v_1 \ot \dots \ot v_{i-1} \ot d v_i \ot v_{i+1} \ot \dots \ot v_n~.~~ 
\raisetag{1.5ex}
	}
For $t^k=v_1 \ot \dots \ot v_n \in T^k(\f)\,,\ k=\sum_{i=1}^n \deg(v_i)\,,$ and 
$\t{t}^l \in T^l(\f)$ we get from \rf[dvv] 
\eq{
d(t^k \ot \t{t}^l)=d(t^k) \ot \t{t}^l + (-1)^k t^k \ot d \t{t}^l~.~~ 
   }
Thus, $d$ defined by \rf[datt] is a graded differential of the tensor algebra 
$T(\f)\,.$ Moreover, $d$ is also a graded differential of the graded Lie 
algebra $T(\f)\,:$ 
\eqa{rcl}{
d[t^k,\t{t}^l] &=& d(t^k \ot \t{t}^l - (-1)^{kl} \t{t}^l \ot t^k) \npb \\ &=& 
(d(t^k) \ot \t{t}^l -(-1)^{(k+1)l} \t{t}^l \ot d t^k ) +(-1)^k (t^k \ot 
d \t{t}^l - (-1)^{k(l+1)} d (\t{t}^l) \ot t^k) \npb \\ 
&=& [dt^k,\t{t}^l] + (-1)^k [t^k,d\t{t}^l] ~.~~
	}
Now, from $d(V(\f) \op V(d\f)) \subset V(\f) \op V(d\f)$ we conclude that 
$d$ is also a graded differential of the graded Lie subalgebra 
$\tW{*} \subset T(\f)\,.$ 

Next, we show that $dI'(\f) \subset I'(\f)\,:$ 
\eqas{rclrcl}{
d(\lambda \delta_a-\delta_{\lambda a}) &=& 
\lambda \delta_{da}-\delta_{d(\lambda a)}~, \qquad{} &
d(\lambda \delta_{da}-\delta_{d(\lambda a)}) &=& 0~, \\ 
d(\delta_{a} +\delta_{\t{a}}-\delta_{a+\t{a}}) &=& \delta_{da} 
+\delta_{d\t{a}}-\delta_{d(a+\t{a})}~, 
\qquad{} & 
d(\delta_{da} +\delta_{d\t{a}}-\delta_{d(a+\t{a})}) &=& 0~,  \\ 
\mc{6}{c}{
\begin{array}{rcl}
d([\delta_{a} ,\delta_{\t{a}}]-\delta_{[a,\t{a}]}) &=& 
[\delta_{da} ,\delta_{\t{a}}] 
+ [\delta_{a} ,\delta_{d\t{a}}] - \delta_{d[a,\t{a}]}~,~~ \\
d([\delta_{da} ,\delta_{\t{a}}]+ [\delta_{a} ,\delta_{d\t{a}}] 
- \delta_{d[a,\t{a}]}) &=&
-[\delta_{da} ,\delta_{d\t{a}}] + [\delta_{da} ,\delta_{d\t{a}}] = 0~. 
\end{array} }
      }
Since $d(V(\f) \op V(d\f)) \subset V(\f) \op V(d\f)\,,$ we get from \rf[iip] 
\eq{
dI(\f) \subset I(\f)~.~~ 
   }
Therefore, the graded differential $d$ on $\tW{*}$ induces a graded 
differential on $\W{*}$ denoted by the same symbol: 
\eq{
d( \vpi+I(\f)):= d \vpi + I(\f)~, \quad  \vpi \in \tW{*} ~.
   }
Hence, $(\W{*}\,,\,[~,~]\,,\,d)$ is a graded differential Lie algebra. 

We extend the involution $*: a \mapsto -a$ on $\f$ to an involution of the free 
vector spaces $V(\f)$ and $V(d\f)$ by 
\al{
(\lambda \delta_a)^* &:=-\lambda \delta_{a}~, & (\lambda \delta_{da})^* & 
:=-\lambda \delta_{da}~.~~  \label{invV}
   }
We obtain an involution of $T(\f)$ by 
\eq{
(v_1 \ot v_2 \ot \dots \ot v_n)^*:=v_n^* \ot \dots \ot v_2^* \ot v_1^* ~,
   }
fulfilling 
\eq[tts]{
(t \ot \t{t})^*= \t{t}^* \ot t^*~.~~ 
	}
Formula \rf[tts] induces the following property of the Lie bracket \rf[comt]: 
\eq{
[t^k,\t{t}^l]^*=-(-1)^{kl} [{t^k}^*,\t{t}^l{}^*]~.~~ 
   }
Because of $(V(\f) \op V(d\f))^* = V(\f) \op V(d\f)$ we get an involution 
on $\tW{*}$ by restricting the involution on $T(V)$ to its graded Lie 
subalgebra $\tW{*}\,.$ Obviously, we have ${I'(\f)}^* = I'(\f)\,,$ giving 
$I(\f)^* = I(\f)\,.$ Therefore, we obtain an involution on $\W{*}$ by 
\eq{ 
(\vpi+I(\f))^*:=\vpi^*+I(\f)~, \quad \vpi \in\tW{*}~.
   }

\subsubsection{The Universality Property of $\W{*}$}

The graded differential Lie algebra $\W{*}$ is universal in the following 
sense: 
\begin{prp} 
\label{upr}
Let $\Lambda^*\f=\bigoplus_{n \in \N_0} \Lambda^n \f$ be an $\N$--graded Lie 
algebra with graded differential $\mathrm{d}: \Lambda^n\f \to \Lambda^{n+1}\f$ 
such that 
\\ 
$\mathrm{i)} \qquad  \Lambda^0\f=\pi(\f)\,,$ for a surjective homomorphism 
$\pi:\f \to 
\pi(\f)$ of Lie algebras,
\\            
$\mathrm{ii)} \qquad \Lambda ^*\f$ is generated by $\pi(\f)$ and 
$\mathrm{d}\pi(\f)$ as the set of repeated commutators. 
\\ 
Then there exists a differential ideal $I_{\Lambda} \subset \W{*}$ such that 
$\Lambda^*\f \cong \W{*}/I_{\Lambda}\,.$ 
\end{prp}
\noindent
\textbf{Proof:} We define a surjective mapping $\t{p}:\tW{*} \to \Lambda^*\f$ 
by 
\eqa{rcl}{
\t{p}(\lambda\delta_a) &:=& \pi(\lambda a)~,~~	\npb \\
\t{p}(d \vpi) &:=& \mathrm{d}(\t{p}(\vpi))~,~~ \npb \\
\t{p}([\vpi,\t{\vpi}]) &:=& [\t{p}(\vpi),\t{p}(\t{\vpi})] ~,~~ 
       }
for $a \in g$ $\vpi,\t{\vpi} \in \tW{*}$ and $\lambda \in \R\,.$ Obviously, 
$\t{p}(I(\f))=0\,.$ Therefore, by factorization with respect to $I(\f)$ we get 
a surjection $p:\W{*} \to \Lambda^*\f$ by $p(\vpi+I(\f)):=\t{p}(\vpi)\,,$ for 
$\vpi \in \tW{*}\,.$ We have $p(d \ker p)=0\,,$ therefore, 
$I_{\Lambda}=\ker p$ is the desired differential ideal of $\W{*}\,:$ 
\eq{
\Lambda^*\f \cong \W{*} \,/ \, \ker p~.~~  \tag*{\qedsymbol}
  }
Proposition~\ref{upr} tells us that each graded differential Lie algebra 
generated by $\pi(\f[g])$ and its differential is obtained by factorizing 
$\W{*}$ with respect to a differential ideal. For the setting described by 
an L--cycle, such a differential ideal is canonically given. This leads to a 
canonical graded differential Lie algebra, see Section~\ref{repr}.

\subsubsection{Summary}

To summarize: We have defined a universal graded differential Lie algebra 
$\W{*}=\bigoplus_{n=0}^\infty \W{n}$ over a Lie algebra $\f\,,$ with: 
\vs{-\topsep}
\begin{itemize} 
\item[--] graded commutator $[~,~]: \W{k} \times \W{l} \to \W{k+l}\,,$ 
\vs{-\parsep} \vs{-\itemsep}

\item[--] universal differential $d: \W{k} \to \W{k+1}\,,$ which is linear, 
nilpotent and obeys the graded Leibniz rule.  \vs{-\parsep} \vs{-\itemsep}

\item[--] involution $~*: \W{k} \to \W{k}\,.$ 
\end{itemize} \vs{-\topsep}
Explicitly, we have the following properties: 
\seq{
\label{lie}
\eqa{rl}{
1)~~ & [\omega^k,\t{\omega}^l] =-(-1)^{kl} [\t{\omega}^l, \omega^k] ~,~~ 
\yn \label{liea} \npb \\ 
2)~~ & [\omega^k, \lambda \t{\omega}^l +\t{\lambda} \t{\t{\omega}}^l] 
= \lambda [\omega^k, \t{\omega}^l] + \t{\lambda} 
[\omega^k, \t{\t{\omega}}^l]~,	\label{lieb} \yn \\ 
3)~~ & (-1)^{km} [\omega^k,[\t{\omega}^l,  \t{\t{\omega}}^m]] 
+ (-1)^{lk} [\t{\omega}^l, [ \t{\t{\omega}}^m, \omega^k]] 
+ (-1)^{ml} [\t{\t{\omega}}^m  ,[\omega^k,\t{\omega}^l]]=0 ~, 
\hs{3em} \yn \label{liec} \\ 
4)~~ & d[\omega^k,\t{\omega}^l] =[d\omega^k,\t{\omega}^l] 
+(-1)^k [\omega^k,d\t{\omega}^l]~,~~ \yn \label{dder} \\ 
5)~~ & d^2 \omega^k =0~,~~ \yn \npb \\ 
6)~~ & [\omega^k,\t{\omega}^l]^* =-(-1)^{kl} [\omega^k{}^*,\t{\omega}^l{}^*] 
{} ~,~~ \yn \label{invu} 
     }}
for $\omega^k \in \W{k}\,,~\t{\omega}^l, \t{\t{\omega}}^l \in \W{l}\,,~ 
\t{\t{\omega}}^m \in \W{m}$ and $\lambda, \t{\lambda} \in \R\,.$ 
 
\subsubsection{A Canonical Representation of Elements of $\W{*}$}

It is convenient to fix a canonical ordering in elements of $\W{k}\,,\ k \geq 
1\,.$ First, let 
\al{ 
\iota(a) &:=\delta_{a}+I(\f)~, & \iota(da) &:= \delta_{da}+I(\f)~,
   }
for $a \in \f\,.$ The first equation establishes an isomorphism 
$\W{0} \cong \f\,.$ We shall represent elements $\omega^1 \in \W{1}$ as 
\eqa{rcl}{ 
\omega^1 &=& \iota(d\t{a})+\tsum_{\alpha,z \geq 1} [\iota(a_{\alpha}^z), 
[ \dots [\iota(a_{\alpha}^2), [\iota(a_{\alpha}^1), \iota(d a_{\alpha}^0)]]  
\dots ]] \\ 
&\equiv& \tsum_{\alpha,z \geq 0} [\iota(a_{\alpha}^z), 
[ \dots [\iota(a_{\alpha}^2), [\iota(a_{\alpha}^1), \iota(d a_{\alpha}^0)]] 
\dots ]] ~,~~ \label{udla}
	}
where $\t{a},a_{\alpha}^i \in \f$ and the sums are finite. To avoid possible 
misunderstandings concerning this notation we fix throughout this thesis the 
following convention: Beginning with $z=1\,,$ the index $\alpha$ first runs 
from 1 to $\alpha_1>0$ and labels the terms \smallskip 

$[\iota(a^1_1),\iota(da^0_1)], \dots , 
[\iota(a^1_{\alpha_1}),\iota(da^0_{\alpha_1})]$ 
\smallskip 
\\ 
in \rf[udla]. Then, for $z=2\,,$ the index $\alpha$ runs from $\alpha_1+1$ to 
$\alpha_2 > \alpha_1$ and labels the commutators \smallskip 

$[\iota(a^2_{\alpha_1+1}), [\iota(a^1_{\alpha_1+1}),\iota(da^0_{\alpha_1+1})]], 
\dots , [\iota(a^2_{\alpha_2}),[\iota(a^1_{\alpha_2}),\iota(da^0_{\alpha_2})]]$ 
\smallskip 
\\
in \rf[udla], and so on. 
Therefore, the pair $(i,\beta)$ of indices labelling an element 
$a^i_{\beta} \in \f$ does never occur more than once in the sum \rf[udla]. 
Moreover, we identify the term belonging to the pair $(\alpha=0,z=0)$ of 
indices with $\iota(d\t{a})\,,$ as already indicated in \rf[udla]. 

Now, we write down elements $\omega^k \in \W{k}\,,~k \geq 2\,,$ recursively as 
\eq{ 
\omega^k=\tsum_{\alpha} [\omega^1_{\alpha},\t{\omega}^{k-1}_{\alpha}]~, \quad
\omega_{\alpha}^1 \in \W{1}~,~~\t{\omega}_{\alpha}^{k-1} \in \W{k-1}~,~~ 
\mbox{finite sum}~.~~ \label{on} 
   }
There are two things to check concerning \rf[on]. First, for 
$\t{\omega}^n \equiv \sum_{\alpha} [\t{\omega}^1_{\alpha} , 
\t{\t{\omega}}^{n-1}_{\alpha}] \in \W{n} \,,$ with 
$\t{\omega}^1_{\alpha} \in \W{1}$ and $\t{\t{\omega}}^{n-1}_{\alpha} 
\in \W{n-1} \,,$ we must show that also $[\omega^0,\t{\omega}^n] \in \W{n}$ can 
be represented in the standard form \rf[on], for any $\omega^0 \in \W{0}\,.$ 
But this follows from the graded Jacobi identity 
\rf[liec]: 
\eqa{rcl}{ 
[\omega^0,\t{\omega}^n] &=&  [\omega^0,\tsum_{\alpha} [\t{\omega}^1_{\alpha}, 
\t{\t{\omega}}^{n-1}_{\alpha}] ] \npb \\
&=& -\tsum_{\alpha} [\t{\omega}^1_{\alpha}, [\t{\t{\omega}}^{n-1}_{\alpha} , 
\omega^0]] - (-1)^{n-1} \tsum_{\alpha} [\t{\t{\omega}}^{n-1}_{\alpha}, 
[\omega^0,\t{\omega}^1_{\alpha}]] \npb \\ 
&=& \tsum_{\alpha} \big( [\t{\omega}^1_{\alpha}, 
[ \omega^0, \t{\t{\omega}}^{n-1}_{\alpha} ]] 
+ [[\omega^0,\t{\omega}^1_{\alpha}], \t{\t{\omega}}^{n-1}_{\alpha} ] \big) ~. 
	 }
Second, we must show that the commutator $[\omega^k,\t{\omega}^l] 
\in \W{k+l}\,,$ for $2 \leq k \leq l\,,$ can be represented in the standard 
form \rf[on] of an element of $\W{k+l}\,,$ provided that both 
$\omega^k \in \W{k}$ and $\t{\omega}^l \in \W{l}$ are written down recursively 
in the form \rf[on]. Using again \rf[lieb] and \rf[liec] we get for 
$\omega^k = \sum_{\alpha} [\omega^1_{\alpha}, \t{\t{\omega}}^{k-1}_{\alpha}]$ 
\eqa{rl}{ 
[\omega^k,\t{\omega}^l] =& -(-1)^{lk} \tsum_{\alpha} 
[\t{\omega}^l,[\omega^1_{\alpha}, \t{\t{\omega}}^{k-1}_{\alpha}]]  \npb \\ 
=& \tsum_{\alpha} \big( [\omega^1_{\alpha}, [ \t{\t{\omega}}^{k-1}_{\alpha} , 
\t{\omega}^l ]] +(-1)^k [\t{\t{\omega}}^{k-1}_{\alpha},\
[\omega^1_{\alpha},\t{\omega}^l]] \big)~. 
	}
Repeating this calculation for the commutators 
$[ \t{\t{\omega}}^{k-1}_{\alpha}, \t{\omega}^l]$ and 
$[\t{\t{\omega}}^{k-1}_{\alpha},[\omega^1_{\alpha},\t{\omega}^l]]\,,$ we can 
recursively decrease the degree $k$ until we arrive at degree $1\,.$ 
 
Now we can easily prove 
\eq[selu]{ 
(\omega^k)^* = -(-1)^{k(k-1)/2} \omega^k~, \quad \omega^k \in \W{k}~.
	 }
By definition, \rf[selu] holds for $k=0\,.$ From \rf[udla] and \rf[invu] we get 
for $\omega^1 \in \W{1}$ 
\eqa{rcl}{ 
{\omega^1}^* &=& \tsum_{\alpha,z \geq 0} [\iota(a_{\alpha}^z), [\dots , 
[\iota(a_{\alpha}^1), \iota(d a_{\alpha}^0)]  \dots ]]^* \npb \\  
&=& \tsum_{\alpha,z \geq 0} [\iota(a_{\alpha}^z), 
([ \dots, [\iota(a_{\alpha}^1), \iota(d a_{\alpha}^0)]	\dots ])^*] \npb \\ 
&=&  \dots = \tsum_{\alpha,z \geq 0} [\iota(a_{\alpha}^z),[ \dots , 
[\iota(a_{\alpha}^1), (\iota(d a_{\alpha}^0))^*]  \dots ]] = - \omega^1~. 
	 }
In the same way we get from \rf[on] and \rf[invu] for $\omega^k \in \W{k}$ 
\eqa{rcl}{ 
{\omega^k}^* &=& \tsum_{\alpha} [\omega^1_{\alpha}, 
\t{\omega}^{k-1}_{\alpha}]^*=(-1)^{k-1} \tsum_{\alpha} 
[\omega^1_{\alpha},(\t{\omega}^{k-1}_{\alpha})^*] 
= (-1)^{(\sum_{i=2}^{k-1} i)} \omega^k \\ 
&=& -(-1)^{k(k-1)/2} \omega^k~.~~ 
	 }

\subsection{The Graded Differential Lie Algebra $\Omega^*_D \mathfrak{g}$} 
\label{repr} 
 
\subsubsection{Definition of $\P{*}$}

Following the procedure for K--cycles we define an involutive 
representation\footnote{the symbol $\pi$ is already used but there 
is no danger of confusion} $\pi$ of the universal differential Lie algebra 
$\W{*}$ introduced in Section~\ref{tudla} in the graded Lie algebra $\B(h)$ of 
bounded operators on $h\,,$ where $h$ is the Hilbert space of the L--cycle 
given in Definition \ref{lc}. We underline that $\pi$ will not be a 
representation of graded Lie algebras with differential. The definition of 
$\pi$ uses almost the whole input contained in the L--cycle. First, using the 
grading operator $\Gamma,$ we define a $\Z_2$--grading structure on the vector 
space $\mathscr{O}(h)$ of linear operators on the Hilbert space $h\,,$ 
$\mathscr{O}(h)=\mathscr{O}_0(h) \op \mathscr{O}_1(h)\,,$ by 
\al{ 
\mathscr{O}_0(h) \Gamma &= \Gamma \mathscr{O}_0(h)~, &
\mathscr{O}_1(h) \Gamma &= -\Gamma \mathscr{O}_1(h)~.
   }
This enables us to introduce the graded commutator for $\Z_2$--graded 
linear operators on $h\,:$ If $A_i \in \mathscr{O}_i(h)$ and 
$B_j \in \mathscr{O}_j(h) \cap \B(h)$ we define
\eq{
[ A_i ,B_j]_g := A_i \circ B_j -(-1)^{ij} B_j \circ A_i \equiv 
-(-1)^{ij} [ B_j ,A_i]_g ~,~~
   }
on the subset $h'=\mathrm{domain}(A_i) \cap \{\bsj \in h~,~~ B_j \bsj \in 
\mathrm{domain}(A_i)\}$ of $h\,.$ In certain cases it may be possible to 
extend $h'\,.$ One has $A_j \in \B(h)$ iff $h'=h\,.$ 

Let us define a linear mapping $\t{\pi}: \tW{*} \to \B(h)$ by 
\seq{
\label{pid}
\eqa{rcl}{
\t{\pi}(\lambda\delta_a) &:=& \pi(\lambda a)~,~~ \yn \label{pi} \npb \\ 
\t{\pi}( \lambda\delta_{da}) &:=& [-\iu D,\pi(\lambda a)]_g \equiv 
[-\iu D , \pi(\lambda a)]~,~~ \yn \label{pie} \npb \\ 
\t{\pi}([\vpi^k,\t{\vpi}^l]) &:=& [\t{\pi}(\vpi^k),\t{\pi}(\t{\vpi}^l)]_g ~, 
\hs{3em}  \yn \label{pif} 
	 }}
for $a \in \f\,,\ \vpi^k \in \tW{k}\,,$ $\t{\vpi}^l \in \tW{l}$ and 
$\lambda \in \R\,.$ Note that $\pi(a)$ and $[D, \pi(a)]$ are bounded due to 
Definition~\ref{lc} so that the r.h.s.~of equations \rf[pi] and \rf[pie] belong 
to $\B(h)\,.$ Now, due to $\pi(\f) \subset \mathscr{O}_0(h)$ and $D \in 
\mathscr{O}_1(h)\,,$ we get from \rf[pid] 
\al{
\label{b01}
\t{\pi}(\tW{2k}) &\subset \mathscr{O}_0(h)~, & \t{\pi}(\tW{2k+1}) &\subset
\mathscr{O}_1(h)~.
	}
	 
Next, we show that $\t{\pi}:\tW{*} \to \B(h)$ is an involutive representation, 
where we recall that the involution in $\B(h)$ is defined as 
usual by means of the scalar product $\langle~,~\rangle_h$ on $h\,:$ 
\eq[Binv]{ 
\langle \bsj,\tau^* \t{\bsj} \rangle_h:=\langle \tau \bsj,\t{\bsj} \rangle_h~,
\quad \forall \bsj, \t{\bsj} \in h~,~~ \tau \in \B(h)~.~~
	 }
First, from \rf[invV], \rf[pi] and the fact that $\pi:\f \to \B(h)$ is an 
involutive representation we get 
\[ 
\t{\pi}((\lambda\delta_a)^*)=-\t{\pi}(\lambda\delta_{a})= -\pi(\lambda a) 
= (\pi(\lambda a))^* = (\t{\pi}(\lambda\delta_a))^* ~.~~ 
\] 
Second, from \rf[invV], \rf[pie] and the selfadjointness of $D$ we obtain 
\eqa{rcl}{
\pi((\lambda \delta_{da})^*) &=& -\pi(\lambda\delta_{da}) 
=\iu (D \circ \pi(\lambda a) -\pi(\lambda a) \circ D) \npb \\ 
&=& -(-\iu)^* (D^* \circ (\pi(\lambda a))^*-(\pi(\lambda a))^* \circ D^*) 
\npb \\
&=& -\{-\iu (\pi(\lambda a) \circ D -D \circ \pi(\lambda a))\}^*  
= (\pi(\lambda \delta_{da}))^*~. 
	 }
Now we get by induction that $\t{\pi}$ is an involutive representation on 
$\tW{*}\,.$ 
 
Observe that 
\eq{ 
\t{\pi}(I(\f)) \equiv 0~.~~ 
   }
Therefore, the involutive representation $\t{\pi}:\tW{*} \to \B(h)$ induces an 
involutive representation $\pi:\W{*} \to \B(h)$ by 
\eq{ 
\pi(\vpi+I(\f)):=\t{\pi}(\vpi)~, \quad \vpi \in \tW{*}~.
   }

\subsubsection{Definition of $\D{*}$}

In the same way as for K--cycles there may exist $\omega \in \W{*}\,,$ 
fulfilling $\pi(\omega)=0$ but not $\pi(d \omega)=0\,.$ Therefore, $\P{*}$ is 
not a differential Lie algebra. But there is a canonical construction 
towards such an object. Let us define 
\al{
\mathcal{J}^*\!\f[g] &= \ker\pi + d \ker \pi = \bigoplus_{k=0}^\infty 
\mathcal{J}^k\!\f[g]~, & \mathcal{J}^k\!\f[g] &= \mathcal{J}^*\!\f[g] 
\cap \W{k} ~. \label{Zet}
	}
To obtain a differential Lie algebra we first prove 
\begin{lem} 
$\mathcal{J}^*\!\f[g] $ is a graded differential ideal of the graded Lie 
algebra $\W{*}\,.$  \label{ideal} 
\end{lem} 
\noindent
\Proof{ It is clear that $\ker \pi$ is an ideal of $\W{*}\,.$ Then, for 
$j^k \in \ker \pi \cap \W{k}$ and $\omega \in \W{*}$ we have, see \rf[dder],   
\eqa{rcl}{ 
[dj^k , \omega]= d([j^k,\omega])-(-1)^k [j^k ,d \omega]~. 
	 }
Because of $[j^k,d\omega] \in \ker \pi$ and $d([j^k,\omega]) 
\in d \ker \pi \,,$ $\mathcal{J}^*\!\f[g] $ is an ideal of $\W{*}\,.$ Moreover, 
it is obviously a differential ideal: $d \mathcal{J}^*\!\f[g] \subset 
\mathcal{J}^*\!\f[g] \,,$ due to $d^2=0\,. $ 
}
By virtue of Proposition~\ref{upr}, the canonical differential ideal \rf[Zet] 
gives rise to a graded differential Lie algebra $\D{*}\,:$ 
\seq{
\al{
\D{*} &= \bigoplus_{k=0}^\infty \D{k}~,& \D{k} &:=\W{k}/\mathcal{J}^k\!\f[g]~.
   }
There is a canonical isomorphism 
\eq{
\dfrac{\W{k}}{\mathcal{J}^k\!\f[g]} \cong \dfrac{\W{k} / 
(\ker \pi \cap \W{k})}{\mathcal{J}^k\!\f[g] / (\ker \pi \cap \W{k})}~,~~ 
       }
establishing the isomorphism
\eq{
\D{k} \cong \P{k}/\J{k}~.~~ 
       }
In particular, one has 
\al{
\D{0} &\cong \P{0} \equiv \pi(\f)~,& \D{1} &\cong \P{1}~.
   }
   }
Let $\vrs$ denote the projection onto equivalence classes, $\vrs: \P{k} \to 
\D{k}\,.$ In this notation, the commutator and the differential on $\D{*}$ 
are defined as 
\seq{
\eqa{rcl}{ 
[\vrs \circ \pi(\omega^k),\vrs \circ \pi(\t{\omega}^l)] &:=& 
\vrs ([\pi(\omega^k),\pi(\t{\omega}^l)]_g) ~,~~ \label{defcom} \yn \npb \\ \yn 
d(\vrs \circ \pi(\omega^k)) &:=& \vrs \circ\pi(d\omega^k)~,~~ \label{defdif} 
	 }
for $\omega^k \in \W{k}$ and $\t{\omega}^l \in \W{l}\,.$ From \rf[defcom] there 
follows that $\D{*}$ is a graded Lie algebra, and the bracket 
$[~,~] : \D{*} \times \D{*} \to \D{*}$ has properties analogous to \rf[lie]. 
For $\vrr^k=\vrs \circ \pi(\omega^k)$ 
and $\t{\vrr}^l=\vrs \circ \pi(\t{\omega}^l)$ we have with \rf[defcom] and 
\rf[defdif] 
\eqa{rcl}{
d [\vrr^k,\t{\vrr}^l] &=& \vrs \circ \pi(d[\omega^k,\t{\omega}^l])= 
\vrs \circ \pi([d\omega^k,\t{\omega}^l]+(-1)^k [\omega^k,d\t{\omega}^l]) 
\npb \\
&=& [d\vrr^k,\t{\vrr}^l] + (-1)^k [\vrr^k,d\t{\vrr}^l]~.~~ \label{dtt} \yn
	 }
         }
Obviously, $d^2 \equiv0$ on $\D{*}\,.$ This means that $d$ is a graded 
differential on $\D{}\,.$ Moreover, we have 
\eq{ 
(\vrs \circ \pi(\omega^k))^* =\vrs \circ \pi((\omega^k)^*) ~, \quad 
\omega^k \in \D{k}~,
   }
because $\pi$ is an involutive representation and $\J{*}$ is invariant under 
the involution. From \rf[selu] we get 
\eq[seld]{ 
{\vrr^n}^* = -(-1)^{n(n-1)/2} \vrr^n~, \quad  \vrr^n \in \D{n} ~.
	 }

\subsection{Towards the Analysis of the Differential Ideal} 
\label{tad}
 
Our goal is the analysis of the ideal $\J{*}\,.$ For this purpose we 
define 
\eqa{rl}{ 
\sigma \big( \sum_{\alpha,z \geq 0} [ \iota(a_{\alpha}^z),[ \dots 
[\iota(a_{\alpha}^1), &
\iota(d a_{\alpha}^0)] \dots ]]\big) \npb \\[-1.5ex] 
:= & \sum_{\alpha} [\pi(a_{\alpha}^z),[ \dots [\pi(a_{\alpha}^1), 
[D^2,\pi(a_{\alpha}^0)]] \dots ]]~,   \label{sig1} \yn
	 }
where $a_{\alpha}^i \in \f[g]\,.$ In particular, from \rf[sig1] we get 
\al{
\sigma(\iota(da)) &= [D^2,\pi(a)]~,& \sigma([\iota(a),\omega^1]) & 
= [\pi(a),\sigma(\omega^1)]~, \label{sig2}
	 }
for $a \in \f$ and $\omega^1 \in \W{1}\,.$ We extend $\sigma$ to $\W{*}\,,$ 
putting $\sigma (\W{0}) \equiv 0$ and 
\eq[swa]{
\sigma(\sum_{\alpha} [\omega^k_{\alpha},\t{\omega}^l_{\alpha}]) 
:= \sum_{\alpha} \big( [\sigma(\omega^k_{\alpha}) ,
\pi(\t{\omega}^l_{\alpha})]_g 
+(-1)^k [\pi(\omega^k_{\alpha}) , \sigma(\t{\omega}^l_{\alpha})]_g \big) ~,~~  
	}
for $\omega^k_{\alpha} \in \W{k}$ and $\t{\omega}^l_{\alpha} \in \W{l}\,.$ Note 
that $\sigma(\omega^k) \in \mathscr{O}_{z_{k+1}}(h)$ if $\pi(\omega^k) \in 
\mathscr{O}_{z_k}(h)\,,$ where $z_n = n \mod 2\,.$ We do not necessarily have 
$\sigma (\omega^k) \in \B(h)\,.$ Now we prove: 

\begin{prp} 
We have $\pi(d\omega^k)=[-\iu D , \pi(\omega^k)]_g + \sigma (\omega^k)\,,$ for 
$\omega^k \in 
\W{k}\,.$ \label{pds} 
\end{prp} 
\noindent
\textbf{Proof:} The Proposition is clearly true for $k=0\,.$ To prove the 
Proposition for $k=1$ we first consider the case $\omega^1=\iota(d a) 
\in \W{1}\,.$ Then we have \smallskip

$[-\iu D,\pi(\omega^1)]_g = [-\iu D, [-\iu D ,\pi(a)]_g]_g 
= [(-\iu D)^2 , \pi(a)] =-\sigma (\iota(d a))\,,$ \smallskip
\\
so that $\pi(d\omega^1)=0\,.$ But this is consistent with 
$d\omega^1=d^2(\iota(a))=0\,.$ Now we prove the Proposition for $k=1$ by 
induction. Because of \rf[sig2], the linearity of $\pi$ and the 
structure of elements of $\W{1}\,,$ see \rf[udla], it suffices to assume 
that the Proposition is true for all $\omega^1 \in \W{1}$ and to show that from 
this assumption there follows \smallskip
 
$\pi(d[\iota(a),\omega^1])=[-\iu D , \pi([\iota(a),\omega^1])]_g 
+ \sigma([\iota(a),\omega^1])\,,$ 
\smallskip
\\ 
for all $a \in \f\,.$ We calculate 
\eqa{rcl}{
\pi(d[\iota(a),\omega^1]) &=& [\pi(\iota(da)),\pi(\omega^1)]_g 
+ [\pi(\iota(a)),\pi(d\omega^1)]_g \npb \\ 
&=& [[-\iu D, \pi(a)]_g,\pi(\omega^1)]_g +[\pi(a),[-\iu D, \pi(\omega^1)]_g 
+ \sigma(\omega^1)]_g \npb \\
&=& [-\iu D, [\pi(a),\pi(\omega^1)]_g]_g + \sigma([\iota(a),\omega^1]) \npb \\
&=& [-\iu D, \pi([\iota(a),\omega^1]) ]_g + \sigma([\iota(a),\omega^1])~. 
	 }

Finally, we extend the proof to any $k$ by induction. For that purpose let us 
assume that the Proposition holds for $k-1\,.$ Due to linearity we can restrict 
ourselves to elements $\omega^k= [\omega^1,\t{\omega}^{k-1}] \in \W{k}\,.$ 
Using \rf[swa] and the graded Jacobi identity we calculate
\al{
\pi(d[ \omega^1, & \t{\omega}^{k-1}]) 
= [\pi(d\omega^1),\pi(\t{\omega}^{k-1})]_g 
- [\pi(\omega^1),\pi(d\t{\omega}^{k-1})]_g \notag \\ 
=& [[-\iu D, \pi(\omega^1)]_g + \sigma(\omega^1), \pi(\t{\omega}^{k-1})]_g 
-[\pi(\omega^1),[-\iu D, \pi(\t{\omega}^{k-1})]_g + \sigma(\t{\omega}^{k-1})]_g 
\notag \\
=& -[ \pi(\t{\omega}^{k-1}), [-\iu D, \pi(\omega^1)]_g ]_g \notag \\
& -(-1)^k [\pi(\omega^1),[\pi(\t{\omega}^{k-1}),-\iu D]_g ]_g 
+ \sigma([\omega^1,\t{\omega}^{k-1}]) \notag \\
=& [-\iu D, [\pi(\omega^1),\pi(\t{\omega}^{k-1})]_g ]_g 
+ \sigma([\omega^1,\t{\omega}^{k-1}]) ~.
\tag*{\qedsymbol}
      }
We recall that 
\eq[js0]{ 
\J{k}= \{ \pi(d\omega^{k-1})~,~~ \omega^{k-1} \in \W{k-1} \cap \ker \pi~\}~.~~ 
	}
From Proposition~\ref{pds} we get the following equivalent characterization: 
\eq[js]{								      
\J{k}= \{ \sigma(\omega^{k-1})~,~~ \omega^{k-1} \in \W{k-1} \cap \ker \pi ~\}~.  
       }
Obviously, $\sigma(\omega^{k-1})$ is bounded if $\pi(\omega^{k-1})=0\,.$ Of 
course, \rf[js] is only a rewriting of \rf[js0], but it is a convenient 
starting point for the analysis of $\J{*}\,.$ 
 
\subsection{Graded Lie Homomorphisms}
\label{glh}

In this subsection we provide the framework for the formulation of connections 
and gauge transformations. 

\subsubsection{Definition of $\H{*}$ and $\hH{*}$}

Let 
\eqa{rl}{
\H{n} := \{ ~& \eta^n \in \mathscr{O}_{z_n}(h)~,~~ z_n=n \mod 2~,~~
\eta^n{}^* =-(-1)^{n(n-1)/2}\, \eta^n ~,~~ \\
& [\eta^n,\P{k}]_g \subset \P{k+n}~,~~ [\eta^n,\J{k}]_g \subset \J{k+n}~\} ~~{}
\label{HHn} \yn 
	  }
be the set of graded Lie homomorphisms of $\P{*}$ of $n^{\text{th}}$ degree. 
Note that $\H{n}$ may contain unbounded operators $\eta$ on $h\,,$ but such 
that 
\[
h'=\mathrm{domain}(\eta) \cap \{\bsj \in h~,~~ \P{*} \bsj \subset 
\mathrm{domain}(\eta)\}
\]
is dense in $h\,.$ This is necessary to ensure that 
the sequence $\{~ [\eta,\pi(\omega)]_g \bsj_n ~\}_n$ of elements of $h\,,$ for 
$\bsj_n \in h'$ and any $\omega \in \W{*}\,,$ converges to 
$\pi(\t{\omega}) \bsj$ if $\bsj_n$ tends to $\bsj \in h\,,$ where 
$\pi(\t{\omega}) \in \P{*}$ is independent of $\bsj_n\,.$ Let 
\eq[cc]{
\tcc{n} :=\{ j^n \in \H{n}~,~~ [j^n,\P{*}]_g =0~\}  
       }
be the graded center of $\P{*}$ of $n^{\text{th}}$ degree. Then, the factor 
space 
\seq{
\label{okto}
\al{
\tH{*} &:= \bigoplus_{n \in \N_0} \tH{n} ~,& \tH{n} &:=\H{n}\,/\, \tcc{n} ~,
   }
is a graded Lie algebra, with the graded commutator given by	
\eqa{rl}{
[[\eta^k + \tcc{k},\t{\eta}^l & + \tcc{l}]_g , \pi(\omega^n)]_g  \\ & := 
[\eta^k,[\t{\eta}^l , \pi(\omega^n)]_g]_g -(-1)^{kl} [\t{\eta}^l, 
[\eta^k,\pi(\omega^n) ]_g ]_g~, \hs*{3em} \yn \npb \label{cck}
   }
   }
for $\eta^k \in \H{k}\,,\ \t{\eta}^l \in \H{l}$ and $\omega^n \in \W{n}\,.$ 
It is easy to check that this equation is well--defined. Obviously, 
$\P{*}$ is a graded Lie subalgebra of $\tH{*}\,.$ 

It is clear that the graded ideal $\J{*}$ of $\P{*}$ is also a graded ideal of 
$\tH{*}\,,$ see \rf[HHn]. Therefore, 
\seq{
\al{
\hH{*} &:= \bigoplus_{n \in \N_0} \H{n} ~,& \hH{n} &:=\tH{n}\,/\, \J{n} ~, 
\label{hH} 
   }
is a graded Lie algebra. Moreover, it is a graded differential Lie algebra, 
too, where the graded differential is defined by 
\eqa{rl}{
[d(\eta^k + \J{k}), & \pi(\omega^n) + \J{n}]_g \yn \npb \label{dhk} \\
&:= \pi \circ d \circ \pi^{-1} ( [\eta^k,\pi(\omega^n)]) 
- (-1)^k [\eta^k, \pi(d\omega^n)]_g + \J{k+n+1}~,
      }
      }
for $\eta^k \in \tH{k}$ and $\omega^n \in \W{n}\,.$ It is obvious that this 
equation is well--defined and that $\D{*}$ is a graded Lie subalgebra of 
$\hH{*}\,.$ Of course, an equivalent characterization of $\hH{n}$ is 
\al{
\label{hHcJ}
\hH{n} &= \H{n} \,/\, \cJ{n}~, & \cJ{n} &= \tcc{n} + \J{n}~.
   }

\subsubsection{The Exponential Mapping}

Let 
\eqa{rl}{
\mathbbm{u}(\f):=\{~& \eta^0 \in \H{0} \cap \B(h)~,~~  \yn \label{bbmu} \npb \\
& \sigma \circ \pi^{-1}([\eta^0,\pi(\omega^k)]_g) - [\eta^0,\sigma(\omega^k)]_g 
\in \J{k+1}~,~~ \forall  \omega^k \in \W{k} ~\}~.
	}
Obviously, $\pi(\f) \subset \mathbbm{u}(\f)\,.$ Let $\cmcal{O}_0 \subset 
\mathbbm{u}(\f)$ be an open neighbourhood of the zero element of 
$\mathbbm{u}(\f)$ and $\cmcal{O}_1 \subset \B(h)$ be an open neighbourhood of 
$\one_{\B(h)}\,.$ For an appropriate choice of $\cmcal{O}_0$ and $\cmcal{O}_1$ 
we define the exponential mapping 
\al{ 
\exp: \cmcal{O}_0 & \to \cmcal{O}_1~, & \exp(\eta) &:= \one_{\B(h)} 
+ \sum_{k=1}^\infty \frac{1}{k!} (\eta)^k~,~~ \eta \in \cmcal{O}_0~.~~ 
   }
The Baker--Campbell--Hausdorff formula for $\eta_{\alpha},\eta_{\beta} \in 
\cmcal{O}_0\,,$ 
\eqa{l}{
\exp(\eta_{\alpha}) \exp(\eta_{\beta})=\exp(\eta_{\gamma})~,~~ \npb \yn \\ 
\eta_{\gamma}=\eta_{\alpha}+\eta_{\beta}+\tfrac{1}{2} 
[\eta_{\alpha},\eta_{\beta}]+\tfrac{1}{12} 
([\eta_{\alpha},[\eta_{\alpha},\eta_{\beta}]] 
-[\eta_{\beta},[\eta_{\alpha},\eta_{\beta}]]) + \dots \in \mathbbm{u}(\f)~,~~ 
       }
implies that we have a multiplication in $\exp(\cmcal{O}_0)\,.$ In particular, 
for $\eta_{\beta}$ proportional to $\eta_{\alpha}$ we get 
\eq[r1r2]{
\exp(\lambda_1 \eta) \exp(\lambda_2 \eta)= \exp((\lambda_1+\lambda_2)\eta) = 
\exp(\lambda_2 \eta) \exp(\lambda_1 \eta)~,~~  
	 }
for $\lambda_1,\lambda_2 \in \R$ and $\eta \in \cmcal{O}_0\,.$ Thus, 
$\exp(\eta)$ is invertible in $\B(h)$ for each $\eta \in \cmcal{O}_0\,,$ and 
the inverse is given by 
\eq{
(\exp(\eta))^{-1}= \exp(-\eta)=\exp(\eta^*)= (\exp(\eta))^*~.~~ 
   }
Therefore, all elements $\exp(\eta)$ are unitary. Since $\B(h)$ is a 
$C^*$--algebra we conclude that for all $\eta \in \mathbbm{u}(\f)$ we have 
\eq[norm]{
\|\exp(\eta)\|=\|\exp(\eta)^* \exp(\eta)\|^{1/2}= \|\one_{\B(h)}\|^{1/2} =1~.~~  
	 }
Hence, our construction leads to the subgroup 
\eq[pex]{
\exp(\mathbbm{u}(\f)):=\{~ {\ts \prod_{\alpha=1}^N } \exp(\eta_{\alpha}) ~,~~ 
\eta_{\alpha} \in \cmcal{O}_0~,~~ N \mbox{ finite }~\}	
	}
of the group of unitary elements of $\B(h)\,.$ 

For $A$ being a linear operator on $h$ and $\eta \in \cmcal{O}_0$ we have 
\eq[aBa]{
\exp(\eta) A \exp(-\eta) = A+ \sum_{k=1}^\infty \dfrac{1}{k !} 
[\underbrace{\eta,\lbrack \eta, \dots , \lbrack \eta}_k , A \rbrack \dots 
\rbrack ] ~.
	}
For $A=\pi(a) \in \pi(\f)$ and $\exp(\eta) =u \in \cmcal{O}_1$ we get 
$u \pi(a) u^* \in \pi(\f)\,.$ For $A=-\iu D$ we get 
$u[-\iu D,u^*] = -\iu (u D u^* - D) \equiv u d (u^*) \in \hH{1}\,,$ because 
with \rf[bbmu] and \rf[dhk] we have 
\eqa{rcl}{
[[{-} \iu D,\eta] , \pi(\omega^k)]_g {+} \J{k+1} &=& 
[-\iu D,[\eta ,\pi(\omega^k)]]_g - [\eta, [ {-} \iu D,\pi(\omega^k)]_g] 
{+} \J{k+1} \npb \\ 
&=& \pi \circ d \circ \pi^{-1}([\eta,\pi(\omega^k)]) 
- \sigma \circ \pi^{-1} ([\eta,\pi(\omega^k)]) \npb \\
&& - [\eta, \pi(d \omega^k)] \!+\! [\eta, \sigma(\omega^k)] + \J{k+1} \npb \\
&=& [d\eta, \pi(\omega^k)]_g + \J{k+1}~.
	}
If $\pi(\omega^k) \in \J{k}$ then $[[-\iu D, \eta] , \pi(\omega^k)]_g \in 
\J{k+1}\,.$ Therefore, there is a natural degree--preserving representation 
$\Ad{\!}$ of $\exp(\pi(\f))$ in $\D{*}$ defined by 
\eqas[Adu]{rcl}{ 
\Ad{u} \pi(a) &:=& u \pi(a) u^* ~, \\
\Ad{u} [-\iu D,\pi(a)] &:=& [-\iu D, \Ad{u} \pi(a)] + [u[-\iu D,u^*], 
\Ad{u} \pi(a)]~, \\ 
\Ad{u} (\pi(\omega^k) + \J{k}) &:=& ( \Ad{u}\pi(\omega^k)) + \J{k}~, \\
\Ad{u} [\vrr,\t{\vrr}] &:=& [\Ad{u} \vrr, \Ad{u} \t{\vrr}]~, 
       }
for $u \in \exp(\H{0})\,,\ a \in \f[g]\,,\ \omega^k \in \W{k}$ and 
$\vrr,\t{\vrr} \in \D{*}\,.$ Note that due to \rf[aBa] we have 
$\Ad{u} \J{k} \subset \J{k}\,.$ 

\subsection{Connections and Gauge Transformations} 
\label{secc} 
 
In this subsection we define the notion of a connection, of its curvature, of 
gauge transformations and of bosonic and fermionic actions. 

\subsubsection{Connection and Curvature}

\begin{dfn}
\label{connect}
A connection on an L--cycle is a pair $(\nabla,\nabla_h)\,,$ where 
\eqa{ll}{
\mathrm{i)}     & \nabla_h: h \to h \mbox{ is linear, odd and skew--adjoint, } 
		  \npb \\ &  \nabla_h \in \mathscr{O}_1(h)~, \quad
                  \langle \bsj,\nabla_h \t{\bsj} \rangle_h 
		  =-\langle \nabla_h \bsj,\t{\bsj} \rangle_h~, \quad
		  \forall \bsj,\t{\bsj} \in h\,, \npb \\
\mathrm{ii)}    & \nabla: \D{n} \to \D{n+1} \mbox{ is linear,} \npb  \\
\mathrm{iii)~~} & \nabla (\pi(\omega^n) + \J{n}) = [\nabla_h, \pi(\omega^n)]_g 
		  + \sigma(\omega^n) + \J{n+1}~, \quad \omega^n \in \W{n}~.
	 }
The operator $\nabla^2: \D{n} \to \D{n+2}$ is called the curvature of the 
connection. 
\end{dfn} 
\noindent
As a consequence of iii) we get with \rf[swa]
\eq{
\nabla ([\vrr^k,\t{\vrr}^l])=[\nabla (\vrr^k),\t{\vrr}^l]+(-1)^k 
[\vrr^k,\nabla (\t{\vrr})^l]~, \quad\vrr^k \in \D{k}~,~~\t{\vrr}^l \in \D{l}~.
   }
\begin{prp}
Any connection has the form $\big( \nabla=d+[\t{\rho},~.~]_g\,,\, 
\nabla_h=-\iu D+\rho \big)\,,$ for $\rho \in \H{1}$ and $\t{\rho} 
:= \rho + \tcc{1} \in \hH{1}\,.$ Its curvature is $\nabla^2=[\theta,~.~]\,,$ 
with $\theta=d\t{\rho}+\th [\t{\rho},\t{\rho}]_g \in \hH{2}\,.$ 
\label{ltheta}
\end{prp}
\noindent
\textbf{Proof:} There is a canonical connection given by 
$(\nabla=d, \nabla_h=-\iu D)\,.$ Items i) and ii) of Definition~\ref{connect} 
are obvious. For iii) we find with Proposition~\ref{pds}
\eq{
[-\iu D,\pi(\omega^k)]_g+ \sigma(\omega^k)=\pi(d \omega^k)~.~~ 
  }
Taking $\omega \in \ker \pi\,,$ we see that iii) is well--defined. 
Let $(\nabla^{(1)},\nabla^{(1)}_h)$ and $(\nabla^{(2)},\nabla^{(2)}_h)$ be two 
connections. Then we get from iii) of Definition~\ref{connect} 
\eq{
(\nabla^{(1)} {-} \nabla^{(2)})(\pi(\omega^k) + \J{k})=
[\nabla^{(1)}_h {-} \nabla^{(2)}_h, \pi(\omega^k)]_g + \J{k+1}~, 
\label{rpj}
   }
for $\omega^k \in \W{k}\,.$  
Now, item ii) yields $\rho := \nabla^{(1)}_h - \nabla^{(2)}_h \in \H{1}\,.$ 
Since a modification of $\rho$ by an element of $\tcc{1}\equiv \cJ{1}$ does not 
change formula \rf[rpj], we get $\nabla^{(1)}-\nabla^{(2)} 
= [\t{\rho}\,,~.~]\,,$ where $\t{\rho} := \rho+\tcc{1} \in \tH{1} 
\equiv \hH{1}\,.$ Taking $(\nabla^{(2)}, \nabla^{(2)}_h)=(d,-\iu D)$ we obtain 
$(\nabla^{(1)},\nabla^{(1)}_h) = \mbox{$(d+[\t{\rho},~.~]_g\,,\, 
-\iu D+\rho)$}\,.$ 

Note that if $\sigma(\omega^k) \subset \J{k+1}$ for all $\omega^k \in \P{k}$ 
then there is $-\iu D \in \H{1}\,.$ Thus, the assertion remains true although 
the connection $(\nabla=d, \nabla_h=-\iu D)$ is not distinguished in this 
case. 

Finally, we compute the curvature $\nabla^2\,.$ For $\omega^k \in \W{k}$ we 
have with \rf[okto]
\al{
\nabla^2 (\pi(\omega^k) & + \J{k}) = \nabla( \pi(d \omega^k) 
+ [\t{\rho},\pi(\omega^k)]_g + \J{k+1}) \notag \\ 
&= [\t{\rho}, \pi(d \omega^k)]_g 
+ \pi \circ d \circ \pi^{-1}([\t{\rho},\pi(\omega^k)]_g) + 
[\t{\rho},[\t{\rho},\pi(\omega^k)]_g]_g + \J{k+2} \notag \\
&\equiv [d\t{\rho}+\th [\t{\rho},\t{\rho}]_g ,\pi(\omega^k) + \J{k}]_g 
=: [\theta,\pi(\omega^k) + \J{k}]~. \tag*{\qedsymbol}
       }
Note that the relation between $\rho \in \H{1}$ and $\rho' \in \H{1}$ in 
\rf[rpj],
\[
[\rho, \pi(\omega^k)]_g + \J{k+1} = [\rho', \pi(\omega^k)]_g + \J{k+1}~,
\]
may have more solutions than $\rho'=\rho+\tcc{1}\,.$ However, we shall regard 
$\rho$ and $\rho'$ as different connection forms if $\rho-\rho' \not\in 
\tcc{1}\,.$ Analogously, the determining equation for $\theta' \in \hH{2}\,,$ 
\[
[\theta',\vrr]_g = [\theta,\vrr]_g \quad \mbox{for all } \vrr \in \D{*}~,
\]
may have more solutions than $\theta' =\theta\,.$ However, we shall select 
always the canonical representative 
$\theta=d \t{\rho} + \th [\t{\rho},\t{\rho}]_g$ in the 
curvature form of the connection $\nabla^2\,.$ Often we shall 
denote $\theta \in \hH{2}$ itself instead of $\nabla^2$ the curvature of the 
connection $(\nabla,\nabla_h)\,.$ 

\subsubsection{The Gauge Group}

\begin{dfn} 
The gauge group of the L--cycle is the group $\mathcal{U}(\f):= 
\exp(\mathbbm{u}(\f[g]))$ defined in \textrm{\rf[pex]}. 
Gauge transformations of the connection are given by 
\smallskip
\\
\centerline{ $(\nabla, \nabla_h) \longmapsto (\nabla', \nabla_h') 
:=(\Ad{u} \nabla \Ad{u^*}\,,\, u \nabla_h u^*)\,, \quad 
u \in \mathcal{U}(\f)\,.$} \label{ggroup} 
\vs*{-2ex}
\end{dfn} 
We must check that the definition of gauge transformations of a connection is 
compatible with Definition \ref{connect}: 
\eqa{rl}{
[\nabla_h', & \pi(\omega^n)]_g + \J{n+1} 
= u [\nabla_h , u^* \pi(\omega^n) u ]_g u^* + \J{n+1} \npb \\
&= \Ad{u} \big( \nabla (\Ad{u^*}(\pi(\omega^n)+\J{n})) 
- \sigma(\pi^{-1} \circ \Ad{u^*} \circ \pi(\omega^n)) + \J{n+1} \big) \npb \\
&= \nabla' (\pi(\omega^n)+\J{n}) - \Ad{u} (\sigma(\pi^{-1} \circ \Ad{u^*} 
\circ \pi(\omega^n))) 
+ \J{n+1}~.
	}
Thus, the definition is consistent iff $\sigma(\pi^{-1} \circ \Ad{u} 
\circ \pi (\omega^n)) + \J{n+1} = \Ad{u}(\sigma(\omega^n)) + \J{n+1}\,.$ But 
this equation is satisfied due to \rf[bbmu]. 

The gauge transformation of the connection form $\rho$ occurring in the
connection $\nabla_h=-\iu D + \rho$ is defined by 
\eq{ 
\nabla_h' =: -\iu D + \gamma_u(\rho)~.~~ 
   }
From $\nabla_h' \bsj= u(-\iu D + \rho) u^* \bsj 
= (-\iu D + u [-\iu D,u^*] + u \rho u^* ) \bsj$ one finds 
\eq[gur]{
\gamma_u(\rho)= u d u^* + u \rho u^* ~.~~   
	}
The gauge transformation of the curvature is due to
\[                                                  
(\Ad{u}\nabla \Ad{u^*})^2 (\vrr^k) =\Ad{u} \nabla^2 \Ad{u^*} \vrr^k 
=u[\theta,u^* \vrr^k u] u^* 
\]
given by 
\eq[guq]{
\gamma_u (\theta)=\Ad{u} \theta= u \theta u^*~.~~ 
	}

\subsubsection{Bosonic and Fermionic Actions}

The Dixmier trace provides a canonical scalar product $\langle~,~\rangle$ on 
$\B(h)\,,$ see \cite{ac}. If the L--cycle is $\mathrm{d}^+$--summable 
(see Definition~\ref{d+s}) we define for $\tau,\t{\tau} \in \B(h)$ 
\eq[Dix]{ 
\langle \tau,\t{\tau} \rangle
:=\Tr_{\omega}\,(\tau^* \t{\tau}\, |D|^{-\mathrm{d}})~.~~ 
	}
We assume that in some sense there exists an extension of this formula to 
linear operators on $h$ belonging to $\H{2}$ (recall that $\H{2}$ is 
bounded on a dense subset of $h$). 
\begin{dfn}
The bosonic action $S_B$ and the fermionic action $S_F$ of the connection 
$(\nabla,\nabla_h)$ are given by
\seq{
\label{action}
\eqa{l}{
S_B(\nabla) = \langle\theta,\theta \rangle_{\hH{2}}
:= \min_{j^2 \in \cJ{2}} \Tr_{\omega} 
((\theta_0 + j^2)^2\, |D|^{-\mathrm{d}}) ~,~~  
\yn \label{actb} \\  \yn
S_F(\bsj,\nabla_h) := \langle \bsj ,\iu \nabla_h \bsj \rangle_h ~, \quad
\bsj \in h~,
    }}
where $\Tr_{\omega}$ is the Dixmier trace, $\langle~,~\rangle_h$ the scalar 
product on $h$ and $\theta_0 \in \H{2}$ any representative of the curvature 
of $\nabla\,.$ 
\end{dfn}
\noindent
Since both $\langle~,~\rangle_{\hH{2}}$ and $\langle~,~\rangle_h$ are invariant 
under unitary transformations \cite{ac} we get from \rf[guq] and Definition 
\ref{ggroup} that the action \rf[action] is invariant under gauge 
transformations 
\eq{
(\nabla,\nabla_h) \longmapsto (\Ad{u} \nabla \Ad{u^*} ,u \nabla_h u^* ) ~, 
\quad \bsj \longmapsto u \bsj~,~~  u \in \mathcal{U}(\f)~.
   }
There is an equivalent formulation of \rf[actb]. Let $\f[e](\theta_0+j^2) \in 
\H{2}$ be those representative of $\theta \in \H{2}\,,$ for which the minimum 
in \rf[actb] is attained. Let $j^2=\tsum_{\alpha} \lambda_{\alpha} 
j^2_{\alpha}\,,$ for $\lambda_{\alpha} \in \R\,,$ be a parameterization of 
$j^2 \in \cJ{2}\,.$ Then, 
\[
0=\frac{d}{d \lambda_{\alpha}}	\Tr_{\omega} 
((\theta_0 + j^2)^2\, |D|^{-\mathrm{d}})= 2\,  
\Tr_{\omega} ((\theta_0 + j^2) j^2_{\alpha})\, |D|^{-\mathrm{d}})~.~~
\]
Thus, $\f[e](\theta_0+j^2) \equiv \f[e](\theta)$ is those representative of 
$\theta\,,$ which is orthogonal to the ideal $\cJ{2}$ with respect to 
$\langle~,~\rangle_{\hH{2}}\,:$ 
\al{
\label{fee}
S_B &= \Tr_{\omega} ((\f[e](\theta))^2\, |D|^{-\mathrm{d}}) ~, &
\Tr_{\omega} (\f[e](\theta)\, \cJ{2}\: |D|^{-\mathrm{d}}) &\equiv 0~.~~ 
	}
The representative $\f[e](\theta)$ is unique, because 
$\Tr_{\omega} (~.~ |D|^{-\mathrm{d}})$ is positive definite \cite{ac}:
\eqa{rcl}{
\Tr_{\omega} (( \f[e](\theta)+j^2)^2\,	|D|^{-\mathrm{d}})
&=& \Tr_{\omega} ((\f[e](\theta))^2\, |D|^{-\mathrm{d}})
+ \Tr_{\omega} ((j^2)^2  |D|^{-\mathrm{d}}) \npb \\
&>& \Tr_{\omega} ((\f[e](\theta))^2\, |D|^{-\mathrm{d}})~, \quad 
\mbox{for $j^2 \neq 0$}~.
 }

\section{L--cycles over Functions $\ot$ Matrix Lie Algebra}
\label{ftm}

\subsection{A Class of L--cycles Relevant to Physics}
\label{nota}

Let $(\f[a],\C^F,\M,\h{\pi},\h{\Gamma})$ be an L--cycle over a matrix Lie 
algebra $\f[a]\,.$ In particular, we have a representation $\h{\pi}$ of 
$\f[a]$ in the Lie algebra $\mat{F}$ of endomorphisms of the Hilbert space 
$\C^F\,.$ Moreover, the grading operator $\h{\Gamma}$ anticommutes with the 
generalized Dirac operator $\M$ and commutes with $\pf\,.$ Both $\M$ and 
$\h{\Gamma}$ belong to $\mat{F}\,.$

Let $X$ be a compact even dimensional Riemannian spin manifold, 
$\dim (X) =N \geq 4\,,$ and let $C^\infty(X)$ be the algebra of real--valued 
smooth functions on $X$. Since $C^\infty(X)$ is a commutative 
algebra, the tensor product
\seq{
\eq{
\f:=C^\infty(X) \ot \f[a]  
   }
over $\R$ is in a natural way a Lie algebra, where the commutator is given by 
\eq{
[f_1 \ot a_1,f_2 \ot a_2] \equiv f_1 f_2 \ot [a_1,a_2]~, \quad
f_1,f_2 \in C^\infty(X)~,~~a_1,a_2 \in \f[a]~.~~  
   }
   }
We introduce the Hilbert space 
\eq{
h:=L^2(X,S) \ot \C^F~,~~
   }
where $L^2(X,S)$ denotes the Hilbert space of square integrable sections of the 
spinor bundle over $X$. The representation $\h{\pi}: \f[a] \to 
\End\,(\C^F)$ and the $C^\infty(X)$--module structure of $L^2(X,S)$ induce a 
natural representation $\pi$ of $\f$ in $\B(h)$: 
\eq[piex]{
\pi(f \ot a)(s \ot \varphi):=fs \ot \h{\pi}(a) \varphi~,~~ 
         }
for $f \in C^\infty(X)\,,~a \in \f[a]\,,~s \in L^2(X,S)$ and $\varphi \in 
\C^F\,.$ We denote by $\ga$ the grading operator and by $\sfD$ the 
classical Dirac operator on the Hilbert space $L^2(X,S)\,,$ see 
Section~\ref{tech} for more details. Then we put 
\seq{
\eqa{rcl}{
D &:=& \sfD \ot \one_F + \ga \ot \M~,~~ \yn  \label{DD} \npb \\
\Gamma &:=& \ga \ot \h{\Gamma}~.~~ \yn	
	 }}
The operator $[D,\pi(f \ot a)]$ is bounded on $h$ for all $f \ot a \in \f\,.$ 
Moreover, $D$ is selfadjoint on $h\,,$ because $\sfD$ and $\ga$ are selfadjoint 
on $L^2(X,S)$ and $\M$ is symmetrical. Next, $\Gamma$ commutes with 
$\pi(\f[g])$ and anticommutes with $D\,.$ Finally, $(\id_h+D^2)^{-1}$ 
is compact, see \cite{g}: The operator $(\id_h+D^2)^{-1}$ is a 
pseudo--differential operator of order $-2$ with compact support and has, 
therefore, an extension to a continuous operator from $H_s$ to $H_{s+2}$ on the 
Sobolev scale $\{H_s\}\,.$ Due to Rellich's lemma, the embedding 
$e: H_t \hookrightarrow H_s$ is compact for $t >s\,.$ Thus, $(\id_h+D^2)^{-1}$ 
considered as 
\[
e `o (\id_h+D^2)^{-1} : H_s \to H_s
\]
is compact, and $(\f,h,D,\pi,\Gamma)$ forms an L--cycle.  

Finally, we brief\/ly sketch how the physical data specified in the 
Introduction fit into this scheme. First, one constructs a Euclidian version of 
the gauge field theory. Now, $X$ is the one--point compactification of the 
Euclidian space--time manifold. The completion of the space of fermions $\bsj$ 
yields the Hilbert space $h$ of the L--cycle. In some cases, it may be 
necessary to work with several copies of the fermions. Given the (Lie) group of 
local gauge transformations $\mathscr{G}\,,$ we take $\f[g]$ as the Lie algebra 
of $\mathscr{G}\,.$ The representation $\pi:\f[g] \to \B(h)$ is just the 
differential $\t{\pi}_*$ of the group representation $\t{\pi}\,.$ The matrix 
$\M$ occurring in the generalized Dirac operator \rf[DD] contains the fermionic 
mass parameters and possibly contributions required by the desired symmetry 
breaking scheme. However, it is necessary that $\ga \ot \M$ coincides with the 
fermionic mass matrix $\widetilde{\M}$ on chiral fermions. The grading 
operator $\Gamma$ represents the chirality properties of the fermions. We 
have $\ga=\g$ in four dimensions. After the Wick rotation to Minkowski space we 
use $\Gamma$ to impose a chirality condition on $h\,.$ 

\subsection{Notations and Techniques}
\label{tech}

\subsubsection{Exterior and Interior Products}

We denote by $\Gamma^\infty(C)$ the set of smooth sections of the 
Clifford bundle $C$ over $X$ and by $C^k \subset \Gamma^\infty(C)$ the set of 
those sections of $C\,,$ whose values at each point $x \in X$ belong to the 
subspace spanned by products of less than or equal $k$ elements of $T^*_xX$ of 
the same parity. In particular, we identify $C^\infty(X) \equiv C^0\,.$ 

We recall \cite{bgv} that there is an isomorphism of vector spaces 
\eq{
c: \Lambda^*(\Gamma^\infty(T^* X)) \to \Gamma^\infty(C)
   }
between $\Gamma^\infty(C)$ and the exterior differential algebra 
$\Lambda^*(\Gamma^\infty(T^* X))$ of antisymmetrized tensor products of the 
vector space of smooth sections of the cotangent bundle over $X.$ In 
particular, the restriction to the first degree yields a vector space 
isomorphism $c: \Gamma^\infty(T^* X) \to C^1\,.$ Therefore, elements 
$c^1 \in C^1$ have the form $c^1=c(\bsw^1)\,,$ 
for $\bsw^1 \in \Gamma^\infty(T^* X)\,.$ 
We use the following sign convention for the defining relation of the Clifford 
action:
\eq{
\th (c(\bsw^1) c(\t{\bsw}^1) + c(\t{\bsw}^1) c(\bsw^1)) \equiv\th \{c(\bsw^1) , 
c(\t{\bsw}^1) \} 
= g^{-1}(\bsw^1 ,\t{\bsw}^1 ) 1 \in C^0~,
   }
where $g^{-1}: \Gamma^\infty(T^* X) \times \Gamma^\infty(T^* X) \to \CX$ is the 
inverse of the metric $g: \Gamma^\infty(T_* X) \times \Gamma^\infty(T_* X) 
\to \CX\,.$ 

Let us define the notion of the exterior product $\wedge$:
\eq[ext]{
c^1_1 \wedge c^1_2 \wedge \dots \wedge c^1_n := \dfrac{1}{n!} 
\sum_{\pi \in P^n} (-1)^{\mathrm{sign}(\pi)} 
c^1_{\pi(1)} c^1_{\pi(2)} \dots c^1_{\pi(n)}~, \quad c^1_i \in C^1~,
       }
where the sum runs over all permutations of the numbers $1, \dots , n\,,$ and 
the product on the r.h.s.\ is pointwise the product in the Clifford algebra. 
Observe that $\wedge$ is associative and that the antisymmetrization \rf[ext] 
yields zero for $n > N=\dim (X)\,.$ 
\begin{dfn}
$\Lambda^n \subset C^n$ is the vector subspace generated by elements of the 
form \rf[ext], with $\Lambda^0 \equiv C^0\,,$ $\Lambda^1 \equiv C^1$ and 
$\Lambda^n \equiv \{0\}$ for $n<0$ and $n > \dim(X)\,.$
\end{dfn}
\noindent
We define the interior product $\ctr: \Lambda^1 \times \Lambda^n \to 
\Lambda^{n-1}$ by 
\seq{
\eqa{rcl}{
c^1_0 \ctr (c^1_1 \wedge c^1_2 \wedge \dots \wedge c^1_n ) &:=& 
\sum_{j=1}^n (-1)^{j+1} \th \{c^1_0 ,c^1_j\} (c^1_1 \wedge \miss{j} 
\wedge c^1_n )~, \hs*{3em}  \label{ctr} \yn
\\[-1ex]
c^1_1 \wedge \miss{j} \wedge c^1_n &:=& c^1_1 \wedge c^1_2 \wedge \dots 
\wedge c^1_{j-1} \wedge c^1_{j+1} \wedge \dots \wedge c^1_n ~.~~  \yn
      }
      }
The interior product \rf[ctr] is extended to $\ctr: \Lambda^k \times 
\Lambda^n \to \Lambda^{n-k}$ by 
\eqa{rl}{
(\t{c}^1_1 \wedge  \t{c}^1_2 \wedge \dots \wedge \t{c}^1_k ) &
\ctr (c^1_1 \wedge c^1_2 \wedge \dots \wedge c^1_n ) \\ & 
:= \t{c}^1_1 \! \ctr ( \dots  \ctr \! (\t{c}^1_{k-1} \! \ctr ( \t{c}^1_k 
\ctr (c^1_1 \wedge c^1_2 \wedge \dots \wedge c^1_n ))) \dots )\,.~~ \yn
   }
\begin{lem} 
For $c^1_i \in C^1$ we have    \label{lemcc}
\seq{
\eqa{rcl}{
\th ( c^1_0(c^1_1 \wedge \dots \wedge c^1_n )+(-1)^n (c^1_1 \wedge \dots \wedge 
c^1_n ) c^1_0 ) 
&=& c^1_0 \wedge c^1_1 \wedge c^1_2 \wedge \dots \wedge c^1_n ~, 
\hs*{3em} \yn  \label{cwcc}  \npb \\
\th ( c^1_0(c^1_1 \wedge \dots \wedge c^1_n ) 
-(-1)^n (c^1_1 \wedge \dots \wedge c^1_n ) c^1_0 ) 
&=& c^1_0 \ctr (c^1_1 \wedge c^1_2 \wedge \dots \wedge c^1_n) ~. 
\hs*{3em} \yn \label{cccc}
    }}
\end{lem} 
\noindent
\Proof{ The assertion is clear for orthogonal bases. }

\subsubsection{Exterior Differential and Codifferential}

Let $\{e^j\}_{j=1}^N$ be an arbitrary selfadjoint basis of 
$\Gamma^\infty(T^*X)$ and $\{e_j\}_{j=1}^N$ its dual basis of 
$\Gamma^\infty(T_* X)\,.$ Duality of $\{e_j\}_{j=1}^N$ and $\{e^j\}_{j=1}^N$ 
is understood in the sense 
\eq{
e^j ( e_i) \equiv \langle e^j, e_i \rangle = \delta^j_i~
   }
and selfadjointness means $c(e^j)=c(e^j)^*\,.$ Let $\nabla_{\!v}$ be the 
Levi--Civita covariant derivative with respect to the vector field 
$v \in \Gamma^\infty(T_*X)\,.$ Then we define the exterior differential 
$\bfd: \Lambda^k \to \Lambda^{k+1}$ on $\Lambda^*$ by
\eq[d]{
\bfd c^k:=\sum_{j=1}^N c(e^j) \wedge  \nabla_{\!e_j} (c^k)~, \quad 
c^k \in \Lambda^k ~.
      }
The proof that $\bfd$ is indeed a graded differential uses the fact that 
the Levi--Civita connection has vanishing torsion, see (with different sign 
conventions) \cite{bgv}. There is a natural scalar product 
$\langle~,~\rangle_{\Lambda^*}$ on $\Lambda^*$:
\eq[ix]{
\langle c^k,\t{c}^l \rangle_{\Lambda^*}:= \int_X \!\! \vg \; 
\tr_c ({c^k}^* \t{c}^l) ~, \quad c^k \in \Lambda^k~,~~ \t{c}^l \in \Lambda^l~,
       }
where $\tr_c:\Gamma^\infty(C) \to \CX$ is pointwise the trace in the Clifford 
algebra and $\vg$ the canonical volume form on $X\,.$
The scalar product \rf[ix] vanishes for $k \neq l\,.$ Via this scalar product 
we define the codifferential $\bfd^*: \Lambda^{k} \to \Lambda^{k-1}$ on 
$\Lambda^*$ as the operator dual to the exterior differential $\bfd$:
\eq{
\langle \bfd c^k,\t{c}^{k+1} \rangle_{\Lambda^*} 
=: \langle c^k,\bfd^* \t{c}^{k+1} \rangle_{\Lambda^*}~, \quad 
\forall c^k \in \Lambda^k~,~~ c^{k+1} \in \Lambda^{k+1}~.~~
   }
\begin{lem}  \label{lemcod}
Within our conventions one has the representation 
\eq[cod]{
\bfd^* c^k = - \sum_{j=1}^N c(e^j) \ctr \nabla_{\!e_j} (c^k)~.~~ 
	} \vs*{-3ex}
\end{lem}
\Proof{ The proof is straightforward. One has to use Lemma~\ref{lemcc}, 
the invariance of the trace under cyclic permutations, the Leibniz rule for 
$\nabla_v$ and the identity $\nabla_v(\vg)\equiv0$ for the Levi--Civita 
connection. 
}
Note that -- in contrast to what its name suggests -- $\bfd^*$ is not a 
derivation. Using \rf[cod] one easily derives for $c^1_i \in C^1 
\equiv\Lambda^1$ the formula
\eqa{rl}{
\bfd^* (c^1_1 & \wedge	c^1_2 \wedge \dots \wedge c^1_n) 
\yn \label{dsc} \npb \\[-1ex] 
=& \tsum_{k=1}^n \big( - (-1)^{k+1} \nabla_{\!g^{-1}(c^{-1}(c^1_k))} 
(c^1_1 \wedge \miss{k} \wedge c^1_n) 
+ (-1)^{k+1} \bfd^* (c^1_k) (c^1_1 \wedge \miss{k} \wedge c^1_n) \big) ~,
	}
where $g^{-1}$ is treated as an isomorphism from $\Gamma^\infty(T^*X)$ to 
$\Gamma^\infty(T_*X)\,.$ 

\subsubsection{Identities for the Dirac Operator}

In terms of the above introduced selfadjoint bases $\{e^j\}_{j=1}^N$ of 
$\Gamma^\infty(T^*X)$ and $\{e_j\}_{j=1}^N$ of $\Gamma^\infty(T_* X)\,,$ the 
classical Dirac operator is given by \cite{bgv}
\eq[sfd]{
\sfD=\sum_{j=1}^N \iu c(e^j) \nabla^S_{\!e_j}~.~~ 
	}
Here, $\nabla^S_v$ is the Clifford covariant derivative on $L^2(X,S)$ 
with respect to the vector field $v\,.$ It has the property 
\eq[nsv]{
[\nabla^S_v,c(\bsw)]= c(\nabla_v \bsw) \equiv \nabla_v c(\bsw) ~,~~ 
	}
for any differential form $\bsw\,.$ With \rf[d] this gives immediately
\eq[sfdf]{
[\sfD,f]=\tsum_{j=1}^N \iu c(e^j) [\nabla^S_{\!e_j},f] \equiv \iu \bfd f 
\equiv \iu c(\sfd f)~, \quad f \in \CX~,~~ 
	 }
where $\sfd$ is the usual exterior differential on the exterior differential 
algebra. The grading operator on $L^2(X,S)$ is $\ga=-\iu^{N/2} c(\vg)\,,$ 
fulfilling
\eqa{rcl}{
\sfD\ga + \ga \sfD &=& \iu^{-1+N/2} \tsum_{j=1}^N ( c(e^j) [\nabla^S_{\!e_j}, 
c(\vg)] + (c(e^j) c(\vg) + c(\vg) c(e^j) ) \nabla^S_{\!e_j}) \npb \\
&=& \iu^{-1+N/2} \tsum_{j=1}^N ( c(e^j) c(\nabla_{\!e^j}(\vg)) + 
2 c(e^j) \wedge c(\vg)	\nabla^S_{\!e_j}) \equiv0~,~~  \yn
	 }
because of the properties $\nabla_{\!v}(\vg)\equiv0$ and $c(e^j) \wedge c(\vg) 
\in \Lambda^{N+1} \equiv0\,.$ Therefore, the Dirac operator $\sfD$ is an odd 
first order differential operator. One has $\ga^2=c(\vg) c(\vg) 
=\det g^{-1}\,.$ If we restrict ourselves to an orthogonal metric, which we do 
for the rest of this work, then we have $\ga^2=1\,.$ 

Next, using \rf[d], \rf[cod] and Lemma \ref{lemcc} we have for 
$c^k \in \Lambda^k$
\eqa{rcl}{
({-}\iu \sfD) c^k -(-1)^k c^k (-\iu \sfD) &=& \tsum_{j=1}^N ( c(e^j) 
[\nabla^S_{\!e_j}, c^k] + (c(e^j) c^k -(-1)^k c^k c(e^j) ) \nabla^S_{\!e_j}) 
\npb \\
&=& \bfd c^k - \bfd^* c^k + 2 \tsum_{j=1}^N c(e^j) \ctr c^k\, \nabla^S_{\!e_j} 
\yn \label{dds} \npb
\\
&=& \bfd c^k - \bfd^* c^k + 2 \tsum_{i=1}^k (-1)^{i+1} c^1_1 \wedge \miss{i} 
\wedge c^1_k \, \nabla^S_{\!g^{-1}(c^{-1}(c^1_i))} ~,
        }
if $c^k=c^1_1 \wedge c^1_2 \wedge \dots c^1_k\,,$ $c^1_i \in \Lambda^1\,.$ The 
last identity in \rf[dds] is due to 
\eqa{rcl}{
2 \tsum_{j=1}^N c(e^j) \ctr c^k\, \nabla^S_{\!e^j} 
&=& \tsum_{j=1}^N \tsum_{i=1}^k (-1)^{i+1} \{c(e^j),c^1_i\} \, c^1_1 \wedge 
\miss{i} \wedge c^1_k \, \nabla^S_{\!e^j} \npb \\
&=& 2 \tsum_{j=1}^N \tsum_{i=1}^k (-1)^{i+1} g^{-1}(e^j ,c^{-1}(c^1_i)) \, 
c^1_1 \wedge \miss{i} \wedge c^1_k \, \nabla^S_{\!e_j} \npb \\
&=& 2 \tsum_{i=1}^k (-1)^{i+1} c^1_1 \wedge \miss{i} \wedge c^1_k \, 
\nabla^S_{\!g^{-1}(c^{-1}(c^1_i))} ~.
	 }
In particular, 
\eq[lapl]{
[\sfD^2,f]= \Delta f - 2 \nabla^S_{\!\grad f}~, \quad f \in \CX~, 
	 } 
where $\grad f:=g^{-1} (\sfd f)$ is the vector field dual to $\sfd f$ and 
$\Delta$ the scalar Laplacian,
\eq{
\Delta f\equiv \bfd^* \bfd f =- \tsum_{i,j=1}^N g^{-1}(e^i,e^j) 
(\nabla_{\! e_i} \nabla_{\! e_j}-\nabla_{\nabla_{e_i} e_j}) (f) ~.~~  
   }

\subsection{The Representation of $\W{*}$ on the Hilbert Space} 
\label{da}

\subsubsection{Decomposition of the Matrix Lie Algebra}

For physical applications we are interested in the case that the matrix Lie 
algebra $\f[a]$ decomposes into 
\eq[aaa]{
\f[a]=\f[a]' \op \f[a]''~,~~ 
	}
Here, $\f[a]'$ is unitary and semisimple, i.e.\ a direct sum of simple unitary 
Lie algebras, and $\f[a]''$ is a direct sum of copies of the Abelian Lie 
algebra $\u1\,,$ each of them represented in the form 
$\u1_{(i)}=\R \mathrm{b}_{(i)}\,.$ In particular, direct sum means that 
elements of different direct sum subspaces always commute. For each copy of 
$\u1\,,$ the representation $\h{\pi}(\mathrm{b})$ shall have the following 
property: There exist $\lambda^z \in \R$ such that 
\eq[eez]{
[\h{\pi}(\mathrm{b}),\M]=\tsum_{z \geq 2} \lambda^z 
[\underbrace{\h{\pi}(\mathrm{b}), \lbrack \dots \lbrack \h{\pi}(\mathrm{b}), 
\lbrack \h{\pi}(\mathrm{b}) }_z,\M \rbrack \rbrack \dots \rbrack ]~.~~ 
	}
For simplicity, we restrict ourselves to the case $\f[a]''=\u1\,,$ where 
\rf[eez] is given by 
\eqas[ee]{rcll}{
[\h{\pi}(\mathrm{b}),[\h{\pi}(\mathrm{b}),[\h{\pi}(\mathrm{b}), \M]]] 
&=&& [\h{\pi}(\mathrm{b}),\M] \qquad \mbox{or}	\quad{} \\{}
[\h{\pi}(\mathrm{b}),[\h{\pi}(\mathrm{b}),[\h{\pi}(\mathrm{b}), \M]]] 
&=& - &[\h{\pi}(\mathrm{b}),\M] ~.~
	}
The extension to the general case is obvious.

\subsubsection{The Construction of $\P{1}$}

Our goal is to construct the graded differential Lie algebra 
$\D{*}$ associated to the L--cycle $(\f,h,D,\pi,\Gamma)\,,$ see 
Section~\ref{repr}. For this purpose we first have to construct the graded Lie 
algebra $\P{*}$ associated to this L--cycle. We denote by $\p{*}$ the 
corresponding graded Lie algebra associated to the L--cycle 
$(\f[a],\C^F,\M,\h{\pi},\h{\Gamma}) \,.$ From \rf[sfdf] we get
\eq[Df]{
[D,\pi(f \ot a)]=\iu \bfd f \ot \h{\pi}(a) + f\ga \ot [\M,\h{\pi}(a)]~, \quad 
a \in\f[a]~,~~f \in \CX~,
       }
where $\bfd$ is the exterior differential \rf[d]. Using that $C^0$ is an 
Abelian algebra, that elements of $C^0$ commute with elements of 
$C^1$ and that $\h{\pi}$ is a representation we obtain for elements of 
$\P{1}\,,$ see \rf[udla] and \rf[pid], 
\seq{
\label{saz}
\eqa{rl}{
\pi \big( \tsum_{\alpha,z\geq 0} & [\iota(f^z_{\alpha} \ot a^z_{\alpha}), 
[ \dots [\iota(f^1_{\alpha} \ot a^1_{\alpha}), 
\iota(d(f^0_{\alpha} \ot a^0_{\alpha}))] \dots ]] \big) \\  
& = \tsum_{\alpha,z \geq 0} [\pi(f^z_{\alpha} \ot a^z_{\alpha}), 
[ \dots [\pi(f^1_{\alpha} \ot a^1_{\alpha}), 
[-\iu D,\pi(f^0_{\alpha} \ot a^0_{\alpha})]] \dots ]] \\
& = \tsum_{\alpha,z \geq 0} f^z_{\alpha} \cdots f^1_{\alpha} \bfd f^0_{\alpha} 
\ot \h{\pi} ([a^z_{\alpha}, [ \dots [a^1_{\alpha}, a^0_{\alpha}] \dots ]]) 
\hs*{3em} \label{saza} \yn \\ 
& + \tsum_{\alpha, z \geq 0} f^z_{\alpha} \cdots f^1_{\alpha} f^0_{\alpha} \ga 
\ot \h{\pi}([\iota(a^z_{\alpha}), [ \dots [\iota(a^1_{\alpha}), 
\iota(da^0_{\alpha})] \dots ]]) ~. \hs*{3em} \yn \label{sazb} 
      }}
Here, we have $f^j_{\alpha} \in C^0\,,~a^j_{\alpha} \in \f[a]\,,$ and $d$ 
denotes the universal differential on both the universal differential Lie 
algebras over $\f$ and $\f[a]\,;$ it is clear from the context on which of 
them. The same notational simplification was used for the factorization 
mappings $\iota\,.$ There are two different contributions in this formula, 
\rf[saza] belongs to $C^1 \ot \p{0}$ and \rf[sazb] to $C^0 \ga \ot \p{1}\,.$ If 
it was possible to put all $f_{\alpha}^0$ equal to constants without changing 
the range of \rf[sazb] then the lines \rf[saza] and \rf[sazb] would be 
independent. This is possible iff 
\[
f_0^0 \ot \h{\pi}(\iota(d a_0^0)) \subset \tsum_{\alpha, z \geq 1} f^z_{\alpha} 
\cdots f^1_{\alpha} f^0_{\alpha} \ga \ot \h{\pi}([\iota(a^z_{\alpha}), 
[ \dots [\iota(a^1_{\alpha}), \iota(da^0_{\alpha})] \dots ]]) ~.
\]
But this is indeed the case, due to \rf[eez] for $a^0_0 \in \f[a]''$ and the 
fact that $\f[a]'$ is semisimple. Namely, for a semisimple Lie algebra $\f[a]'$ 
we have $[\f[a]',\f[a]'] =\f[a]'\,,$ see \cite{h}. This means that 
\eq[all]{
\forall \, a' \in \f[a]~~ \exists \, a'_{\alpha},\t{a}'_{\alpha} \in \f[a]'~:~~ 
a'=\tsum_{\alpha} [a'_{\alpha}, \t{a}'_{\alpha}]~.~~
	}
Then, $\iota(da')=\tsum_{\alpha} \big( [\iota(a'_{\alpha}), 
\iota(d\t{a}'_{\alpha})] -[\iota(\t{a}'_{\alpha}), \iota(d a'_{\alpha})] 
\big)\,.$ 
Here we see the importance of the restrictions imposed to $\f[a]\,,$ we will 
meet further examples in the sequel. 

Now, from the definition \rf[udla] of $\W[a]{1}$ there follows that \rf[sazb] 
can attain any element of $C^0 \ga \ot \p{1}\,.$ We split elements 
$a^j_{\alpha} \in \f[a]$ according to \rf[aaa]. Since commutators containing 
elements of the Abelian part vanish, there is a non--vanishing contribution of 
elements of $\f[a]''$ to \rf[saza] only from the term 
$\bfd \t{f}^0_0 \ot \h{\pi}(a_0^0)\,,$ for $a_0^0 \in \f[a]''\,.$ 
Therefore, the coefficient of elements of $\pf[a'']$ is the Clifford 
action of a total differential. We denote the space $\bfd C^0 \subset C^1$ by 
$B^1$ (``boundary''). In the case of the semisimple Lie algebra $\f[a]'\,,$ the 
line \rf[saza] attains any element of $C^1 \ot \pf[a']\,,$ due to \rf[all].  
Thus, we get the final result
\eq[p1fin]{
\P{1}=(\Lambda^1 \ot \pf[a']) \op (B^1 \ot \h{\pi} (\f[a]'')) \op 
(\Lambda^0 \ga \ot \p{1})~.~~ 
	  }
This means that elements $\tau^1 \in \P{1}$ are of the form
\eq[tau1]{
\tau^1=\sum_{\alpha} \big( c^1_{\alpha} \ot \h{\pi}(a'_{\alpha})+b^1_{\alpha} 
\ot \h{\pi}(a''_{\alpha})+f_{\alpha} \ga \ot \h{\pi}(\omega^1_{\alpha}) \big)~,
	 }
where $c^1_{\alpha} \in C^1\,,~b^1_{\alpha} \in B^1\,,~f_{\alpha} \in C^0\,,~
a'_{\alpha} \in \f[a]'\,,~a''_{\alpha} \in \f[a]''$ and $\omega^1_{\alpha} 
\in \W[a]{1}\,.$ 

\begin{prp}  
\label{prpog}
\eq[pOg]{
\P{n}=(\Lambda^n \ot \pf[a']) \op \big( \bigoplus_{j=1}^n \Lambda^{n-j} 
\ga^j \ot (\p{j} + \K{j-2}{n}) \big)~, \quad n \geq 2~, \raisetag{1.5ex}
	}
where $\K{j}{n}$ is zero for $j <0\,,$ $n < j+2$ or $n>N+j+2\,.$ For $j \geq 0$ 
and $j+2 \leq n \leq N+2$ it is recursively defined by
\seq{
\eqa{rcl}{
\K{0}{2} &:=& \{\pf,\pf\}~, \qquad \K{0}{N+2} := [\pf, \{\pf,\pf[a']\}] ~,~~ 
\yn \npb \\ 
\K{0}{n} &:=& \{\pf, \pf[a']\}~,~~3 \leq n \leq N +1 ~,~~ \\
\K{j}{n} &:=& \{\pf, \p{j} + \K{j-2}{j+1} \} + [\p{1}, \K{j-1}{j+1}]_g \;, 
\npb \\\multicolumn{3}{r}{2+j \leq n \leq N+j+1\;,~j>0\;, \hs*{3em}} 
\npb \yn \\
\K{j}{N+j+2} &:=& [\pf, \K{j}{j+2}] + [\p{1}, \K{j-1}{N+j+1}]_g ~,~~ j>0~.~~
     }}
\vs*{-2ex}
\end{prp}
\Proof{ The proposition is proved by induction. We need the following 
two identities:
\seq{
\label{id}
\eqa{rl}{
(\t{c}^1 \ot &\h{\pi}(\t{a})) (c^{n-j} \ga^j \ot A^j) -(-1)^n 
(c^{n-j} \ga^j \ot A^j) (\t{c}^1 \ot \h{\pi}(\t{a})) \npb \\ 
&= \th (\t{c}^1 c^{n-j} + (-1)^{n-j} c^{n-j} \t{c}^1 ) \ga^j \ot 
(\h{\pi}(\t{a}) A^j - A^j \h{\pi}(\t{a}) )  \label{id1} \yn \npb \\ 
& + \th (\t{c}^1 c^{n-j} - (-1)^{n-j} c^{n-j} \t{c}^1 ) \ga^j \ot 
(\h{\pi}(\t{a}) A^j + A^j \h{\pi}(\t{a}) )  ~,~~ 
\\[1.5ex]
\multicolumn{2}{l}{
(\t{f} \ga \ot \h{\pi}(\t{\omega}^1)) (c^{n-j} \ga^j \ot A^j) -(-1)^n 
(c^{n-j} \ga^j \ot A^j) (\t{f} \ga \ot \h{\pi}(\t{\omega}^1)) }
\npb \\ 
& = (-1)^{n-j} \t{f} c^{n-j} \ga^{j+1} \ot 
(\h{\pi}(\t{\omega}^1) A^j - (-1)^j A^j \h{\pi}(\t{\omega}^1) )  ~,~~  
\label{id2}  \yn
	 }}
for $\t{c}^1 \in \Lambda^1\,,$ $c^n \in \Lambda^n\,,$ $\t{f} \in \Lambda^0\,,$ 
$\t{a} \in \f[a]\,,$ $\t{\omega}^1 \in \W[a]{1}$ and any $A^j \in \mat{F}\,.$
We shall write \rf[tau1] in the form 
\[
\tau^1=\tsum_{\alpha} \big( c^1_{\alpha} \ot \h{\pi}(a_{\alpha}) 
+ f_{\alpha} \ga \ot \h{\pi}(\omega_{\alpha}) \big) ~, 
\]
where $\tsum_{\alpha} c^1_{\alpha} \ot \h{\pi}(a_{\alpha}) 
\equiv \tsum_{\alpha} \big({c^1_{\alpha}}\!' \ot \h{\pi}(a_{\alpha}') 
+ {c^1_{\alpha}}\!'' \ot \h{\pi}(a_{\alpha}'') \big) \,.$ \smallskip

Using \rf[id1], \rf[id2] and Lemma \ref{lemcc} we obtain from 
\rf[on] the following form of elements $\tau^2 \in \P{2}\,:$ 
\seq{
\eqa{rcl}{
\tau^2 &=& {\sum}_{\alpha} (\tau^1_{\alpha} \t{\tau}^1_{\alpha} 
+ \t{\tau}^1_{\alpha} \tau^1_{\alpha} )  \npb \\ \label{tau2} \yn 
&=& {\sum}_{\alpha,\beta,\gamma}  \big( c^1_{\alpha\beta} \wedge 
\t{c}^1_{\alpha\gamma} \ot [\h{\pi}(a_{\alpha\beta}), 
\h{\pi}(\t{a}_{\alpha\gamma})] +  
f_{\alpha \beta} \t{f}_{\alpha \gamma} \ot [\h{\pi}(\omega^1_{\alpha\beta}), 
\h{\pi}(\t{\omega}^1_{\alpha\gamma})]_g \hs*{3em} \npb \\ 
&&+ \t{f}_{\alpha\gamma} c^1_{\alpha\beta} \ga \ot [\h{\pi} (a_{\alpha\beta}), 
\h{\pi}(\t{\omega}^1_{\alpha\gamma})] 
+ f_{\alpha\beta} \t{c}^1_{\alpha\gamma} \ga \ot 
[\h{\pi}(\t{a}_{\alpha\gamma}), \h{\pi}(\omega^1_{\alpha\beta})]  \big) 
+ \kappa^0~, \npb \\ \yn
\kappa^0 &=& {\sum}_{\alpha,\beta,\gamma} c^1_{\alpha\beta} \ctr 
\t{c}^1_{\alpha\gamma} \ot \{ \h{\pi}(a_{\alpha\beta}), 
\h{\pi}(\t{a}_{\alpha\gamma})\} ~.\hs*{3em} 
       }}
All five occurring different types of tensor products are independent. 
This is due to the fact that for non--vanishing $\t{c}^1 \in \Lambda^1$ and 
$c^n \in \Lambda^n$ the equality $\t{c}^1 \wedge c^n =0$ implies 
$\t{c}^1 \ctr c^n \neq 0$ and $\t{c}^1 \ctr c^n =0$ implies 
$\t{c}^1 \wedge c^n \neq 0\,,$ see Lemma \ref{lemcc}. First, $\kappa^0$ 
attains each element of $\Lambda^0 \ot \{\pf,\pf\}\,.$ Moreover, 
$\tsum_{\alpha} f\t{f} \ot [\h{\pi}(\omega^1_{\alpha}), 
\h{\pi}(\t{\omega}^1_{\alpha})]_g$ gives an arbitrary element of 
$\Lambda^0 \ot \p{2}$ and each term in \rf[tau2] containing $\ga$ an arbitrary 
element of $\Lambda^1 \ga \ot \p{1}\,.$ The only not obvious elements are those 
of the form $[\M,\h{\pi}(a)]\,.$ However, they can be represented by \rf[ee] 
for $a=a''$ and for $a=a'$ due to \rf[all] by 
\eq[mtaa]{
[\M,\h{\pi}(\tsum_{\alpha}[a_{\alpha}',\t{a}_{\alpha}'])]=
\tsum_{\alpha} ([[\M,\h{\pi}(a_{\alpha}')],\h{\pi}(\t{a}_{\alpha}')] + 
[\h{\pi}(a_{\alpha}'),[\M,\h{\pi}(\t{a}_{\alpha}')]])~.~~ 
	 }
Finally, $\tsum_{\alpha,\beta,\gamma} c^1_{\alpha\beta} \wedge 
\t{c}^1_{\alpha\gamma} \ot [\h{\pi} (a'_{\alpha\beta}), 
\h{\pi}(\t{a}'_{\alpha\gamma})]$ represents an arbitrary element of 
$\Lambda^2 \ot \pf[a']\,,$ because possible contributions from $\f[a]''$ are 
cancelled by the commutator. Collecting these results, we arrive at \rf[pOg], 
for $n=2\,.$ For $n>2$ one proceeds by induction, see \cite{phd}. 
}
Thus, the computation of $\P{n}$ is reduced to an iterative multiplication of 
matrices only. 

\subsection{Main Theorem}
\label{mthm}

\subsubsection{Definition of $\h{\sigma}$ and $\hsg$}

To derive the structure of $\D{*}\,,$ we first define in analogy to \rf[sig1] 
\eq[sig1a]{
\h{\sigma}( \! \sum_{\alpha,z \geq 0} [ \iota(a_{\alpha}^z),[ \dots 
[\iota(a_{\alpha}^1), \iota(d a_{\alpha}^0)] \dots ]]) := 
\!\! \sum_{\alpha,z \geq 0} [\h{\pi}(a_{\alpha}^z), 
[ \dots [\h{\pi}(a_{\alpha}^1), [\M^2,\h{\pi}(a_{\alpha}^0)]] \dots ]]\;, 
\raisetag{1.5ex}
	  }
for $a^i_{\alpha} \in \f[a]\,.$ We extend $\h{\sigma}$ to a linear map 
$\hsg:\W{*} \to \Gamma^\infty(C) \ot \mat{F}$ by
\eqas[hsg]{l}{
\hsg(\iota(f \ot a)) := 0~, \qquad \hsg(\iota(d(f \ot a))) 
:= f \ot \h{\sigma}(\iota(da)) ~,~~ \\
\hsg([\omega^k,\t{\omega}^l]):=[\hsg(\omega^k),\pi(\t{\omega}^l)]_g 
+(-1)^k [\pi(\omega^k),\hsg(\t{\omega}^l)]_g~,~~ 
     }
for $f \in \CX\,,\ a \in \f[a]\,,\ \omega^k \in \W{k}$ and 
$\t{\omega}^l \in \W{l}\,.$ 

\begin{thm}
\label{maint}
For $\{\pf[a''], \pf[a'']\} \cap \p{2} = 0$ we have 
\eqa{rcl}{
\J{n} &=& \bigoplus_{j=2}^{n} \Lambda^{n-j} \ga^j \ot (\jj{j} + \tK{j-2}{n}) 
\yn \label{maintj} \npb \\
&+& B^N \ga^{n} \ot (\{\pf, \p{n-N-2}+\K{n-4-N}{n-2}\} \cap \p{n-N}) ~,~~ 
	 }
where $B^N=\bfd \Lambda^{N-1}\,,$ $\tK{0}{n}\equiv \K{0}{n}$ and  
\seq{
\label{maintk}
\eqa{rcl}{
\tK{j}{n} &=& \{\pf,\p{j}+\tK{j-2}{n-1} \} 
+ [\p{1},\tK{j-1}{n-1}]_g \yn  \npb \\ 
&+& \h{\sigma}(\h{\pi}^{-1} (\K{j-1}{j+1} \cap \p{j+1})) ~, \quad
2{+}j \leq n \leq N{+}j{+}1~,~~j>0~,~~ \\
\tK{j}{N+j+2} &=& [\pf, \tK{j}{N+j+1} ] 
+ [\p{1},\tK{j-1}{N+j+1}]_g   \yn \npb \\ 
&+& \h{\sigma}(\h{\pi}^{-1} (\K{j-1}{N+j+1} \cap \p{j+1})) ~, \quad j>0~.
	}}
If $\{\pf[a''], \pf[a'']\} \cap \p{2} \neq 0$ then $\J{3}$ must be replaced by 
\[
\J{3} = \J{3} \restriction_{\rf[maintj]}  
+ B^1 \ot (\{\pf[a''], \pf[a'']\} \cap \p{2}) ~.
\]
\vs*{-4ex}
\end{thm}
\noindent
\textbf{Proof:} The proof consists in deriving a formula for $\sigma(\omega^k)$ 
for a given $\omega^k \in \W{k}\,.$ Taking 
$\omega^k \in \W{k} \cap \ker \pi\,,$ we can derive 
the structure of $\J{k+1}\,,$ see \rf[sig1]. We start with $k=1$ and 
proceed for higher degrees by induction. 

\subsubsection*{Introduction}

We consider the splitting 
\[
\h{\omega}^1=d(\iota(a')+\iota(a''))+\tsum_{\alpha,z \geq 1} 
[\iota(a^z_{\alpha}),[ \dots , [\iota(a^2_{\alpha}), [\iota(a^1_{\alpha}), 
\iota(d a^0_{\alpha})]] \dots ]] \in \W[a]{1}~,~~
\]
for $a' = \tsum_{\beta} [a'_{\beta},\t{a}'_{\beta}] \in \f[a]'$ and 
$a'' \in \f[a]''\,.$ Due to \rf[ee] and \rf[all] we can replace 
$\omega^1_0:=\iota(d(a'+a''))$ by  
\eqa{rcl}{
\h{\omega}^1_0 &=& \pm \tfrac{5}{4} [\iota(\mathrm{b}), 
[\iota(\mathrm{b}),\iota(da'')]] 
- \tfrac{1}{4} [\iota(\mathrm{b}),[\iota(\mathrm{b}),[\iota(\mathrm{b}), 
[\iota(\mathrm{b}),\iota(da'') ]]]] \npb \\ 
&&+ \tsum_{\beta} \big( [\iota(a'_{\beta}),\iota(d\t{a}'_{\beta})] 
- [\iota(\t{a}'_{\beta}), \iota(da'_{\beta})] \big)~.
	}
Here, in the first term the plus sign (minus sign) stands if in \rf[ee] the 
equation with the plus sign (minus sign) is realized. Indeed, we have
\al{
\h{\pi}(\h{\omega}^1_0) &\equiv \h{\pi}(\omega^1_0) ~,& 
\h{\sigma}(\h{\omega}^1_0) &\equiv \h{\sigma}(\omega^1_0)~.~~  \label{sbs}
        }
The first formula is due to \rf[ee] for $a''$ and due to the Jacobi identity 
for $a'\,.$ The $\f[a]'$--part of the second formula in \rf[sbs] follows 
immediately from the Jacobi identity. The proof for the $\f[a]''$--part 
consists of algebraic manipulations of \rf[ee], which are not difficult but 
rather lengthy so that they are not listed in this work. The importance of 
the identities \rf[sbs] is that already elements of $\W[a]{1}\,,$ which do not 
contain terms labelled by $z=0\,,$ are sufficient for the construction of 
$\p{1}$ and $\h{\sigma}(\W[a]{1})\,.$ 

\subsubsection{The Proof for $n=2$}

Using \rf[saz] we can represent elements $\omega^1 \in \W{1}$ as 
\seq{
\eqa{l}{
\omega^1 = \tsum_{\alpha,z \geq 0} [\iota(f^z_{\alpha} \ot a_{\alpha}^z), 
[ \dots [\iota(f^1_{\alpha} \ot a_{\alpha}^1), 
\iota(d (f^0_{\alpha} \ot a_{\alpha}^0))] \dots ]] ~,~~ 
\yn \label{o1} \\
\Rightarrow \qquad \pi(\omega^1) = \tsum_{\alpha,z \geq 0} 
\big( \h{c}^{1,z}_{\alpha} \ot \h{\pi}(\h{a}_{\alpha}^z) 
+ \h{f}^z_{\alpha} \ga \ot \h{\pi}(\h{\omega}^{1,z}_{\alpha}) \big) ~,~~\\
\h{f}^z_{\alpha} = f^z_{\alpha} \cdots f^1_{\alpha} f^0_{\alpha} 
\in \Lambda^0~, \qquad
\h{c}^{1,z}_{\alpha} = f^z_{\alpha} \cdots f^1_{\alpha} \bfd f^0_{\alpha} 
\in \Lambda^1 ~,~~ \yn \\
\h{a}^z_{\alpha} = [a^z_{\alpha},[ \dots [a^1_{\alpha}, 
a^0_{\alpha}] \dots ]] \in \f[a]~,~~
\h{\omega}^{1,z}_{\alpha} = [\iota(a^z_{\alpha}),[ \dots [\iota(a^1_{\alpha}), 
\iota(da^0_{\alpha})] \dots ]] \in \W[a]{1}~,~~
      }}
where $a^i_{\alpha} \in \f[a]$ and $f^i_{\alpha} \in \Lambda^0\,.$ Applying the 
map $\sigma$ to $\omega^1$ in \rf[o1] we get~-- using \rf[lapl] and $D^2 \equiv 
\sfD^2 \ot \one_F + 1 \ot \M^2\,,$ see \rf[DD]~-- 
\seq{
\eqa{rcl}{
\mc{3}{l}{
\sigma(\omega^1) = \tsum_{\alpha,z \geq 0} [f^z_{\alpha} \ot 
\h{\pi} (a_{\alpha}^z), [ \dots [f^1_{\alpha} \ot \h{\pi}(a_{\alpha}^1), 
[D^2, f^0_{\alpha} \ot \h{\pi}(a_{\alpha}^0)]] \dots ]] 
\equiv \tsum_{j=0}^3 s_j\;, } \npb \\  
s_0 &=& \hsg(\omega^1)=\tsum_{\alpha,z \geq 0} f^z_{\alpha} \cdots 
f^1_{\alpha} f^0_{\alpha} \ot [\h{\pi}(a_{\alpha}^z), 
[ \dots  [\h{\pi}(a_{\alpha}^1), [\M^2,\h{\pi}(a_{\alpha}^0)]] 
\dots ]] \;, \hs*{3em}	\yn \label{kap0} 
\\
s_1 &=& \tsum_{\alpha,z \geq 0} f^z_{\alpha} \cdots f^1_{\alpha} 
(\Delta f^0_{\alpha}) \ot \h{\pi}([a_{\alpha}^z,[ \dots  [a_{\alpha}^1, 
a_{\alpha}^0] \dots ]])  ~,~~  \yn
\\
s_2 &=& - 2 \tsum_{\alpha,z \geq 0} \,f^z_{\alpha} \cdots f^1_{\alpha} 
\nabla^S_{\!\grad f^0_{\alpha}} \ot \h{\pi}([a_{\alpha}^z, 
[ \dots [a_{\alpha}^1, a_{\alpha}^0] \dots ]])~, \label{kap2} \yn
\\
s_3 &=& 2 \tsum_{\alpha,z \geq 1} \big( f^z_{\alpha} \cdots 
f^2_{\alpha} \nabla_{\!\grad f^0_{\alpha}} (f^1_{\alpha}) \ot 
[\h{\pi}(a^z_{\alpha}),[\dots [\h{\pi}(a^2_{\alpha}), 
\h{\pi}(a^0_{\alpha}) \h{\pi}(a^1_{\alpha})] \dots ]] \npb \\ 
&+& f^z_{\alpha} \cdots f^3_{\alpha} \nabla_{\!\grad f^0_{\alpha}} 
(f^2_{\alpha}) f^1_{\alpha} \ot [\h{\pi}(a^z_{\alpha}), 
[\dots [\h{\pi}(a^3_{\alpha}), \h{\pi}([a^1_{\alpha},a^0_{\alpha}]) 
\h{\pi}(a^2_{\alpha})] \dots ]] + \dots  \npb \\ 
&+& \nabla_{\!\grad f^0_{\alpha}} (f^z_{\alpha}) f^{z-1}_{\alpha} 
\cdots f^1_{\alpha} \ot \h{\pi}([a^{z-1}_{\alpha}, 
[\dots [a^1_{\alpha},a^0_{\alpha}] \dots ]] ) \h{\pi}(a^z_{\alpha}) \big) ~. 
\yn 
       }}
From properties of covariant derivatives we find 
\[
f^z_{\alpha} \cdots f^1_{\alpha} \nabla^S_{\!\grad f^0_{\alpha}} = 
\nabla^S_{\! f^z_{\alpha} \cdots f^1_{\alpha} 
g^{-1}(\sfd f^0_{\alpha})} = \nabla^S_{\! g^{-1}(f^z_{\alpha} 
\cdots f^1_{\alpha} \sfd f^0_{\alpha})} ~.
\]
Next, using \rf[d] and \rf[cod] one easily shows 
\eq[dels]{
f^z_{\alpha} \cdots f^1_{\alpha} (\Delta f^0_{\alpha}) 
= \bfd^* (f^z_{\alpha} \cdots f^1_{\alpha} \bfd f^0_{\alpha}) 
+\nabla_{\! \grad f^0_{\alpha}} (f^z_{\alpha} \cdots f^1_{\alpha})~.~~
	 }
Then, the sum of $s_3$ and the part of $s_1$ corresponding to the second term 
on the r.h.s.\ of \rf[dels] will be denoted by $\h{s}(\omega^1)\,:$
\eqa{rcl}{
\h{s}(\omega^1) &=& s_3 + \tsum_{\alpha,z \geq 1} 
\nabla_{\! \grad f^0_{\alpha}} (f^z_{\alpha} \cdots f^1_{\alpha}) \ot 
\h{\pi}(\h{a}^z_{\alpha}) 
\yn \label{hkap} \npb \\
&=& \tsum_{\alpha,z \geq 1} \big(\, f^z_{\alpha} \cdots f^2_{\alpha} 
\nabla_{\! \grad f^0_{\alpha}} (f^1_{\alpha}) \ot [\h{\pi}(a^z_{\alpha}), 
[\dots [\h{\pi}(a^2_{\alpha}), \{\h{\pi}(a^0_{\alpha}), 
\h{\pi}(a^1_{\alpha})\} ] \dots ]]  
\npb \\ 
&+& \! f^z_{\alpha} \cdots f^3_{\alpha} \nabla_{\! \grad f^0_{\alpha}} 
(f^2_{\alpha}) f^1_{\alpha} \ot [\h{\pi}(a^z_{\alpha}), 
[\dots [\h{\pi}(a^3_{\alpha}), \{\h{\pi}([a^1_{\alpha},a^0_{\alpha}]), 
\h{\pi}(a^2_{\alpha})\}] \dots ]] {+} \dots \npb \\ 
&+& \nabla_{\grad f^0_{\alpha}} (f^z_{\alpha}) f^{z-1}_{\alpha} \cdots 
f^1_{\alpha} \ot \{ \h{\pi}([a^{z-1}_{\alpha}, 
[\dots [a^1_{\alpha},a^0_{\alpha}] \dots ]]), \h{\pi}(a^z_{\alpha}) \} \big) 
\npb\\ 
&\in& \Lambda^0 \ot \{\pf,\pf\}~.
       }
Observe that the terms labelled by $z=0$ do not occur in \rf[hkap]. Collecting 
the results we find
\eq[sw1]{
\sigma(\omega^1)=\h{s}(\omega^1) + \hsg(\omega^1) + \sum_{\alpha,z \geq 0} 
\big( - 2 \nabla^S_{\! g^{-1}\circ c^{-1} 
(\h{c}^{1,z}_{\alpha})} \ot \h{\pi}(\h{a}^z_{\alpha}) 
+ \bfd^* (\h{c}^{1,z}_{\alpha}) \ot 
\h{\pi}(\h{a}^z_{\alpha}) \big) ~.  \raisetag{1.5ex}
	}

\subsubsection*{The Relation between $\pi(\omega^1)$ and $\sigma(\omega^1)$} 

It is clear that $\h{s}(\omega^1) \in \Lambda^0 \ot \{\pf,\pf\}$ and 
$\hsg (\omega^1) \in \Lambda^0 \ot \h{\sigma}(\W[a]{1})\,,$ the question is to 
which amount they are determined by $\pi(\omega^1)\,.$ To answer this question 
we first consider 
\eq[o1f]{
\omega^1 =  \tsum_{\alpha} \tsum_{A=1}^3 [\iota(\rmtf_{\alpha A} 
\ot \t{a}_{\alpha}), \iota(d(\rmf_{\alpha A} \ot a_{\alpha}))] ~,~~ 
a_{\alpha}, \t{a}_{\alpha} \in \f[a]~,~~ 
	}
where 
\eqa{rclrclrcl}{
\rmtf_{\alpha1} &=& f_{\alpha}~, \qquad{} & \rmtf_{\alpha2} &=& -\th~, 
\qquad{} & \rmtf_{\alpha3} &=& -\th (f_{\alpha})^2~,~~	\\
\rmf_{\alpha1} &=& f_{\alpha} \t{f}_{\alpha}~, \qquad{} & 
\rmf_{\alpha2} &=& (f_{\alpha})^2 \t{f}_{\alpha}~, \qquad{} &
\rmf_{\alpha3} &=& \t{f}_{\alpha}~,~~  
                      }
for $f_{\alpha}, \t{f}_{\alpha} \in \CX\,.$ These functions have the properties 
\seq{
\eqa{l}{
\tsum_{A=1}^3 \rmtf_{\alpha A} \rmf_{\alpha A}=0~, \quad 
\tsum_{A=1}^3 \rmtf_{\alpha A} \bfd(\rmf_{\alpha A})=0~, 
\quad \tsum_{A=1}^3 \bfd (\rmtf_{\alpha A}) \rmf_{\alpha A}=0~, 
\hs*{3em} \yn \label{viff}  \npb  \\
\tsum_{A=1}^3 \nabla_{\! \grad(\rmtf_{\alpha A})} (\rmf_{\alpha A})= 
\t{f}_{\alpha} \nabla_{\! \grad(f_{\alpha})} (f_{\alpha}) 
= \t{f}_{\alpha} g^{-1}(\sfd f_{\alpha},\sfd f_{\alpha})~. 
\hs*{3em} \yn \label{viffb}
       }}
Due to \rf[viff] we have $\pi(\omega^1)=0$ and $\hsg(\omega^1)=0\,,$ but for 
\rf[hkap] we get
\eqa{rcl}{
\h{s}(\omega^1) &\equiv& \tsum_{\alpha} \tsum_{A=1}^3 
\nabla_{\!\grad \rmtf_{\alpha A}} (\rmf_{\alpha A}) \ot 
\{\h{\pi}(\t{a}_{\alpha}), \h{\pi}(a_{\alpha})\} \npb \\
&=& \tsum_{\alpha} \t{f}_{\alpha} \nabla_{\!\grad f_{\alpha}} (f_{\alpha}) \ot 
\{\h{\pi}(\t{a}_{\alpha}), \h{\pi}(a_{\alpha})\} ~.
	}
Therefore, $\h{s}(\omega^1)$ is independent of $\pi(\omega^1)$ and attains each 
element of $\Lambda^0 \ot \{\pf,\pf\} \equiv \Lambda^0 \ot \K{0}{2}\,.$ This is 
because \rf[viffb], for an appropriate choice of $f_{\alpha},\t{f}_{\alpha}\,,$ 
attains each given function on $X$ (using a partition of unity if necessary). 
Now we prove
\begin{lem}
$\hsg (\ker \pi \cap \W{1})
=\Lambda^0 \ot \h{\sigma}(\ker \h{\pi} \cap \W[a]{1}) \equiv \Lambda^0 \ot 
\jj{2}~.~~$
\label{lsfin}
\end{lem}
\Proof[Proof of Lemma \ref{lsfin}:] {We introduce a linear map 
$\h{\pi}_{\f[g]}: \W{*} \to \B(h)$ by 
\eqa{rcl}{
\h{\pi}_{\f[g]}(\iota(f \ot a)) &:=& f \ot \h{\pi}(a)~, \qquad
\h{\pi}_{\f[g]}(\iota(d(f \ot a))) := f \ga \ot [-\iu \M,\h{\pi}(a)]~,~~ \\
\h{\pi}_{\f[g]}([\omega,\t{\omega}]) &:=& [\h{\pi}_{\f[g]}(\omega), 
\h{\pi}_{\f[g]}(\t{\omega})]_g~,~~
       }
for $f \in \CX\,,\ a \in \f[a]\,,\ \omega,\t{\omega} \in \W{*}\,.$ For 
$\omega^1 \in \W{1}$ given by \rf[o1] we have
\eqa{rcl}{
\pi(\omega^1) &=& \tsum_{\alpha,z \geq 0} \big(\h{c}^{1,z}_{\alpha} \ot 
\h{\pi}(\h{a}_{\alpha}^z) + \h{f}^z_{\alpha} \ga \ot 
\h{\pi}(\h{\omega}^{1,z}_{\alpha}) \big) ~,~~\npb \\
\h{\pi}_{\f[g]}(\omega^1) &=& \tsum_{\alpha,z \geq 0} 
\h{f}^z_{\alpha} \ga \ot \h{\pi}(\h{\omega}^{1,z}_{\alpha}) ~,~~ 
\npb \yn \label{hpsg} \\
\hsg(\omega^1) &=& \tsum_{\alpha,z \geq 0} 
\h{f}^z_{\alpha} \ot \h{\sigma}(\h{\omega}^{1,z}_{\alpha}) ~.~~
	}
For $\omega^1 \in \ker \pi$ we have $\tsum_{\alpha,z \geq 0} 
\h{c}^{1,z}_{\alpha} \ot \h{\pi}(\h{a}_{\alpha}^z) =0$ and 
$\tsum_{\alpha,z \geq 0} \h{f}^z_{\alpha} \ga 
\ot \h{\pi}(\h{\omega}^{1,z}_{\alpha})=0\,,$ because $\Lambda^1$ and 
$\Lambda^0$ are independent. But this means
\eq{
(\ker \pi \cap \W{1}) \subset (\ker \h{\pi}_{\f[g]} \cap \W{1}) ~ \Rightarrow ~ 
\hsg(\ker \pi \cap \W{1}) \subset \hsg(\ker \h{\pi}_{\f[g]} \cap \W{1}) ~.
   }
It is intuitively clear from \rf[hpsg] that 
\eq[just]{
\hsg(\ker \h{\pi}_{\f[g]} \cap \W{1}) = \Lambda^0 \ot \h{\sigma}(\ker \h{\pi} 
\cap \W[a]{1}) \equiv \Lambda^0 \ot \jj{2}~,~~	
	 }
see \rf[js]. The justification for \rf[just] gives the formalism of 
skew--tensor products, see \cite{kppw} for the general scheme and \cite{phd} 
for the application to our case. Now, by virtue of \rf[sbs] it suffices to take 
\aln{
\omega^1 &= \tsum_{\alpha} \tsum_{\beta,z \geq 1} \, 
[\iota(1 \ot a^z_{\alpha\beta}), [ \dots , [\iota(1 \ot a^2_{\alpha\beta}), 
[\iota(f_{\alpha} \ot a^1_{\alpha\beta}), \iota(d(1 \ot a^0_{\alpha\beta})) 
]] \dots ]] ~, \\
\intertext{with} 
\h{\omega}^1_{\alpha} &:= \tsum_{\beta,z \geq 1} [\iota(a^z_{\alpha\beta}),[ 
\dots , [\iota(a^2_{\alpha\beta}), [\iota(a^1_{\alpha\beta}), 
\iota(d a^0_{\alpha\beta})]] \dots ]] 
\in \ker \h{\pi} \cap \W[a]{1}\;,~ \forall \alpha\,,
       }
where $f_{\alpha} \in \Lambda^0$ and $a^i_{\alpha\beta} \in \f[a]\,.$ It is 
obvious that $\pi(\omega^1) \equiv 0$ and that 
$\sigma(\omega^1)=\hsg(\omega^1) =\tsum_{\alpha} f_{\alpha} \ot 
\h{\sigma}(\h{\omega}^1_{\alpha})$ attains each element of 
$\Lambda^0 \ot \jj{2}\,.$ 
\vs{-2.2\topsep}}
$~$ \hfill {\small{(Lemma \ref{lsfin})}}
\\[1.5ex]
We define a linear map $\nabla_{\Omega}$ from $\P{*}$ to 
(unbounded) operators on $h\,,$ 
\eqas[naw]{rcl}{
\nabla_{\Omega}(c^{n-j} \ga^j \ot A^j) &:=& 
\nabla^S_{\!c^{n-j}} \ga^j \ot A^j~, \quad n-j >0\;, \hs*{2em} \\
\nabla_{\Omega} (f \ga^n \ot A^n) &:=& 0~, \quad  f \in \CX~,~~
	 }
where $c^{n-j} \in \Lambda^{n-j}$ and $A^j \in \mat{F}\,.$ Here and in the 
sequel a covariant derivative with respect to elements of $\Lambda^n$ is 
understood in the sense 
\eq[nln]{
\nabla_{c^1_1 \wedge c^1_2 \wedge \dots \wedge c^1_n} 
:=\tsum_{l=1}^k (-1)^{l+1} c^1_1 \wedge \miss{l} \wedge c^1_n 
\nabla_{\!g^{-1}\circ c^{-1}(c^1_l)} ~, \quad	c^1_i \in \Lambda^1~,
	}
where $c^{-1} : \Lambda^1 \to \Gamma^\infty (T^*X)$ and 
$g^{-1}: \Gamma^\infty (T^*X) \to \Gamma^\infty (T_*X)$ are isomorphisms.

\subsubsection*{The Final Formula for $\sigma(\W{1})$}

Now, we can express \rf[sw1] in terms of $\pi(\omega^1)\,.$ 
For given $\tau^1 \in \P{1}$ let $\pi^{-1}(\tau^1) \in \W{1}$ be an arbitrary 
but fixed representative and $\omega^1 \in \W{1}$ be any representative. Then, 
the set $\{\sigma(\omega^1)\}$ of all elements $\sigma(\omega^1)$ fulfilling 
the just introduced conditions is 
\eq[sw1t]{
\{\sigma(\omega^1)\} = \Lambda^0 \ot (\K{0}{2} + \jj{2} ) 
+ \hsg (\pi^{-1}(\tau^1)) - 2 \nabla_{\Omega} (\tau^1) + \bfd^* \tau^1 ~.~~  
	 }
Putting $\tau^1=0\,,$ i.e.\ $\omega^1 \in \ker \pi \cap \W{1}\,,$ we obtain 
immediately the assertion of the theorem for $n=2\,.$ 

\subsubsection{The Proof for $n=3$}

Formula \rf[sw1t] is the starting point for the construction of 
$\sigma(\W{n})\,,$ $n \geq 2\,,$ out of \rf[swa]. The result is:
\begin{lem}
\label{lemn}
For given $\tau^n \in \P{n}$ let $\pi^{-1}(\tau^n) \in \W{n}$ be an 
arbitrary but fixed representative and $\omega^n \in \W{n}$ be any 
representative. Then we have for n=2
\al{
\{\sigma(\omega^2)\} &= \Lambda^1 \ot (\K{0}{3} + \jj{2} ) 
+ \Lambda^0 \ga \ot (\tK{1}{3} +\jj{3}) \notag \\ 
&+ \hsg (\pi^{-1}(\tau^2)) - 2 \nabla_{\Omega} (\tau^2) + \bfd^* \tau^2 
- \bfd \big( \tau^2 \! \restriction_{\Lambda^0 \ot \{\pf[a''],\pf[a'']\}} 
\big) ~
\label{sw2t}  \\
\intertext{and for $n \geq 3$}
\{\sigma(\omega^n)\} &= \hsg (\pi^{-1}(\tau^n)) - 2 \nabla_{\Omega} (\tau^n) 
+ \bfd^* \tau^n 
+ \sum_{j=2}^{n+1} \Lambda^{n+1-j} \ga^j \ot (\tK{j-2}{n+1} + \jj{j}) \notag 
\\[-1ex] \label{swnh} 
&- \bfd \big( \tau^n \restriction_{\Lambda^{N-1} \ga^{n+1} \ot \{\pf, 
\p{n-1-N}+\K{n-3-N}{n-1}\} } \big)~.~~
	}
\vs*{-4ex}
\end{lem}
\Proof[Remarks to the proof of Lemma~\ref{lemn}]{ The Lemma is proved by 
induction exploiting formula \rf[swa]. The proof is very technical and too long 
to display in this work. For the details see \cite{phd}. It is clear that the 
proof of Lemma \ref{lemn} finishes the proof of Theorem~\ref{maint}. Here, for 
$n=2\,,$ one has to take into account that for $\{\pf[a''],\pf[a'']\} \cap 
\p{2}=0$ and $\tau^2=0$ we have $\bfd \big(\tau^2 \! \restriction_{\Lambda^0 
\ot \{\pf[a''], \pf[a'']\}}\big)=0\,.$ If $\{\pf[a''],\pf[a'']\} \cap \p{2} 
\neq 0$ then a non--vanishing $\Lambda^0 \ot \{\pf[a''], 
\pf[a'']\}$--part of $\tau^2=0$ can be compensated by $\Lambda^0 \ot \p{2}\,,$ 
giving the contribution $B^1 \ot (\{\pf[a''],\pf[a'']\} \cap 
\p{2})$ to the ideal $\J{3}\,.$ The same argumentation yields the boundary 
terms in the second line of \rf[maintj].}

\subsection{The Structure of $\D{*}\,,$ Commutator and Differential} 
\label{cdcr}

\subsubsection{The Structure of $\D{*}$} 

As an immediate consequence of Theorem~\ref{maint} we find
\begin{cor} \label{cor}
If $\{\pf[a''], \pf[a'']\} \cap \p{2} = 0$ we have for 
$n \geq 2$
\eqa{rcl}{
\D{n} &=& \big(\Lambda^n \ot \pf[a'] \big) \op \big( \Lambda^{n-1} \ga 
\ot \p{1} \big) \op  \npb \\  \yn \label{dnfin} 
& \op & ~ \bigoplus_{j=2}^n \big( \Lambda^{n-j} \ga^j \ot 
\big((\p{j} + \K{j-2}{n}) \mod (\jj{j} + \tK{j-2}{n}) \big) \npb \\ 
&& \mod \delta^j_{n-N} B^N \ga^n \ot (\{\pf,\p{n-N-2} 
+ \K{n-N-4}{n-1}\} \cap \p{n-N}) \big) ~.
	 }
If $\{\pf[a''], \pf[a'']\} \cap \p{2} \neq 0$ then 
$\D{3}$ must be replaced by 
\eq{
\D{3} = \D{3} \restriction_{\rf[dnfin]} \mod 
B^1 \ot (\{\pf[a''], \pf[a'']\} \cap \p{2})  ~.~~  \tag*{\qedsymbol}
   }
\end{cor}
\noindent
Therefore, the construction of $\D{n}$ is reduced to the problem of finding the 
factor space $(\p{j} + \K{j-2}{n})\,/\, (\jj{j} + \tK{j-2}{n})\,.$ Here, 
only the matrix Lie algebra $\f[a]$ plays a r\^{o}le. The influence of the 
$\Lambda^*$--part to $\D{n}$ is almost trivial. 

\subsubsection{The Commutator of Elements of $\D{*}$}

For the sake of an easier notation we restrict ourselves to the case 
$\{\pf[a''], \pf[a'']\} \cap \p{2} = 0$ and $(\{\pf,\p{n-N-2} 
+ \K{n-N-4}{n-1}\} \cap \p{n-N})=0\,.$ If these conditions are not fulfilled 
then there are obvious modifications to $\D{3}$ and $\D{n}\,,$ $n \geq N+2\,,$ 
see Corollary \ref{cor}. 

Due to Corollary \ref{cor} and \rf[tau1] we represent elements 
$\vrr^n \in \D{n}$ as
\seq{
\label{vrrnj}
\eqa{c}{
\vrr^n= \dsum_{\alpha} \dsum_{j=0}^n c^{n-j}_{\alpha} \ga^j 
\ot (\h{\pi}(\omega^j_{\alpha}) + \tJ{j}{n})~,~~  \label{vrrn}	\yn \npb \\
\tJ{j}{n}:= \jj{j}+\tK{j-2}{n}~, \qquad \tJ{0}{n}\equiv0~, \qquad \tJ{1}{n} 
\equiv 0~,
\label{tJ} \yn \\[0.5ex]
\begin{array}[c]{ll}
n\geq 2~:~~& c^{n-j}_{\alpha} \in \Lambda^{n-j}~, 
\qquad \h{\pi}(\omega^0_{\alpha}) \in \pf[a']~,
\qquad \h{\pi}(\omega^j_{\alpha}) \in \p{j} \mbox{ for } j>0~,	\npb \\
n=1~: & c^1_{\alpha} \in \Lambda^1 \mbox{ if }\h{\pi}(\omega^0_{\alpha}) 
\in \pf[a'] ~, \qquad
c^1_{\alpha} \in B^1 \mbox{ if } \h{\pi}(\omega^0_{\alpha}) \in \pf[a''] ~,
\npb \\
& c^0_{\alpha} \in \Lambda^0~, \qquad \h{\pi}(\omega^1_{\alpha}) \in \p{1} ~, 
\npb \\
n=0 ~: & c^0_{\alpha} \in \Lambda^0~, \qquad \h{\pi}(\omega^0_{\alpha}) 
\in \pf[a]~.
\end{array} \yn
       }}
The formula for the graded commutator of elements of $\D{*}$ is very simple, 
\eqa{rl}{
\big[ \dsum_{\alpha} & \dsum_{i=0}^k c^{k-i}_{\alpha} \ga^i \ot 
(\h{\pi}(\omega^i_{\alpha}) + \tJ{i}{k}), 
\dsum_{\beta} \dsum_{j=0}^l \t{c}^{l-j}_{\beta} \ga^j \ot 
(\h{\pi}(\t{\omega}^j_{\beta}) + \tJ{j}{l}) \big]_g ~~
\yn \label{vst} \\[-0.5ex] &
= \dsum_{\alpha,\beta} \sum_{i=0}^k \dsum_{j=0}^l  (-1)^{i(l-j)} \, 
c^{k-i}_{\alpha} \wedge \t{c}^{l-j}_{\beta} \ga^{i+j} \ot 
([\h{\pi}(\omega^i_{\alpha}) , \h{\pi}(\t{\omega}^j_{\alpha}) ]_g 
+ \tJ{i+j}{k+l})~,~~
	}
because if the product between $c^{k-i}_{\alpha}$ and $\t{c}^{l-j}_{\beta}$ is 
not completely antisymmetrized then we get a combination of graded 
anticommutators of elements of $\p{*}$ in the second component of the tensor 
product, which contributes to the ideal $\J{*}\,.$ Thus, the graded commutator 
of elements of $\D{*}$ is given by the combination of the exterior product of 
the $\Lambda^*$--parts and the graded commutator of the $\p{*}$--parts modulo 
$\J{*}\,,$ where a graded sign due to the exchange with $\ga$ must be added. 

\subsubsection{The Differential of Elements of $\D{*}$}

Due to \rf[dds] and \rf[nln] we have for $c^k \in\Lambda^k$
\eq{
(-\iu \sfD) c^k - (-1)^k c^k (-\iu \sfD) = \bfd c^k - \bfd^* c^k 
+ 2 \nabla^S_{c^k}~.~~
   }
We apply Proposition~\ref{pds} and Lemma \ref{lemn} to \rf[vrrn], where we 
introduce $\tau^n:= \tsum_{\alpha} \tsum_{j=0}^n c^{n-j}_{\alpha} \ga^j 
\ot \h{\pi}(\omega^j_{\alpha}) \in \P{n}$ and use \rf[naw] and \rf[hsg]. This 
gives

\eqas{rcl}{
d \vrr^n &\equiv& \pi(d\pi^{-1}(\tau^n)) + \J{n+1} \\
&=& \tsum_{\alpha} \tsum_{j=0}^n \big( 
((-\iu \sfD) c^{n-j}_{\alpha} - (-1)^{n-j} c^{n-j}_{\alpha} (-\iu \sfD)) 
\ga^j \ot \h{\pi}(\omega^j_{\alpha}) \\
&& + (-1)^{n-j} c^{n-j}_{\alpha} \ga^{j+1} \ot 
((-\iu\M) \h{\pi}(\omega^j_{\alpha}) - (-1)^j \h{\pi}(\omega^j_{\alpha}) 
(-\iu\M)) \big) \\
&& + \bfd^* \tau^n - 2 \nabla_{\Omega} (\tau^n) + \hsg(\pi^{-1} (\tau^n)) 
+ \J{n+1} \\
&=& \tsum_{\alpha} \tsum_{j=0}^n \big( 
\bfd c^{n-j}_{\alpha} \ga^j \ot (\h{\pi}(\omega^j_{\alpha}) + \tJ{j}{n+1} ) \\
&& + c^{n-j}_{\alpha} \ga^{j+1} \ot ((-1)^{n-j} 
[-\iu \M, \h{\pi}(\omega^j_{\alpha}) ]_g 
+ \h{\sigma}(\omega^j_{\alpha}) + \tJ{j+1}{n+1} ) \big)~.~~ 
	}
Let us say some words on the terms in \rf[sw2t] and \rf[swnh] containing total 
differentials. In general, for 
\[
\tau^k := c^{k-j} \ga^j \ot \h{\pi}(\h{\kappa}^{j-2}_k) \in \Lambda^{k-j} 
\ga^j \ot \K{j-2}{k} \subset \J{k}
\]
we have $\bfd \tau^k \in \J{k+1}\,.$ This is no longer true for $k=2$ and 
$\h{\pi}(\h{\kappa}^{0}_2)=\{\h{\pi}(a''),\h{\pi}(\t{a}'')\}\,,$ with 
$a'',\t{a}'' \in \f[a]''\,.$ 
However, in this case the differential $\bfd \tau^2$ is eliminated by the 
counter\-term $- \bfd \big(\tau^2 \! \restriction_{\Lambda^0 \ot \{\pf[a''], 
\pf[a'']\}} \big)$ in \rf[sw2t]. An analogous property holds for 
\mbox{$k{-}j=N{-}1\,,$} 
where the terms $\bfd \tau^k$ are cancelled by the differentials in \rf[swnh]. 
Therefore, in the following formula for the differentiation rule on $\D{*}$
one must omit these boundary terms. Then we obtain a simple formula:
\eqa{rl}{
d \vrr^n = \big( & (\bfd \ot \one_F)(\tau^n) + [\ga \ot -\iu\M , \tau^n]_g 
\npb \\
&+ (1 \ot \h{\sigma} \circ \h{\pi}^{-1}) \circ \tau^n \circ 
(\ga \ot \one_F) \big) \mod \J{n+1}~,  \yn \label{dt1}
        }
where $\tau^n \in \P{n}$ is an arbitrary representative of 
$\vrr^n \in \D{n}\,.$ Here, the differential $\bfd$ ignores the grading 
operator $\ga\,,$ i.e. $\bfd (c^k \ga) := (\bfd c^k )\ga\,.$ The non--trivial 
part in this formula is to find the spaces $\tJ{j}{n+1}$ constituting the ideal 
$\J{n+1}\,.$ The differential $\bfd \tau^n\,,$ the graded commutator with 
$\ga \ot -\iu\M$ and even the computation of $(1 \ot \h{\sigma} 
\circ \h{\pi}^{-1}) (\tau^n)$ are not difficult for a concrete example. 

\subsection{Local Connections}
\label{lona}

\subsubsection{Making $\D{*}$ to a $\Lambda^*$--Module}

In the case under consideration, an L--cycle over the tensor product of the 
algebra of functions and a matrix Lie algebra, there exists the notion of 
locality. Our goal is to define a multiplication 
\eqa{l}{
\t{\wedge} : \Lambda^k \times \D{n} \to \D{k+n}~,~~ \label{twe}   \yn  \npb \\
\t{c}^k \t{\wedge} ( \sum_{\alpha} \sum_{j=0}^n \! c^{n-j}_{\alpha} \ga^j \ot 
(\h{\pi}(\omega^j_{\alpha}) {+} \tJ{j}{n})) 
:= \! \sum_{\alpha} \sum_{j=0}^n (\t{c}^k \wedge 
c^{n-j}_{\alpha}) \ga^j \ot (\h{\pi}(\omega^j_{\alpha}) {+} \tJ{j}{k+n})\;,  
       }
see \rf[vrrnj]. However, we clearly have problems to do this on the whole 
differential Lie algebra 
$\D{*}$ due to the existence of the boundary spaces $\Lambda^0 \ot \pf[a'']$ 
in $\D{0}\equiv\pi(\f[g])$ and $B^1 \ot \pf[a'']$ in $\D{1}\equiv\P{1}\,.$ 
These boundary spaces in general do not yield elements of $\D{*}$ when we 
multiply them by elements of $\Lambda^*\,.$ Moreover, there are problems if the 
boundary terms $\delta^j_{n-N} B^N \ga^n \ot (\{\pf, \p{n-N-2} 
+ \K{n-N-4}{n-1}\} \cap \p{n-N})$ and $B^1 \ot (\{\pf[a''], 
\pf[a'']\} \cap \p{2})$ in Corollary~\ref{cor} are present. Therefore, 
formula \rf[twe] is understood to hold on subspaces of $\D{*}\,,$ 
where no collision with boundary terms occurs. Then, the multiplication 
\rf[twe] is associative, 
\eq{
(c^k \wedge \t{c}^l) \t{\wedge} \vrr^n = c^k \t{\wedge} (\t{c}^l \t{\wedge} 
\vrr^n)~,~~
   }
for $\t{c}^k \in \Lambda^k\,,\ \t{\t{c}}{}^l \in \Lambda^l$ and 
$\vrr^n \in \D{n}$ (different from boundary spaces). In particular, 
$\D{n}$ carries a natural $\CX$--module structure, where we omit the 
multiplication symbol $\t{\wedge}$ for simplicity:
\eq[fwf]{
f ( \dsum_{\alpha} \dsum_{j=0}^n c^{n-j}_{\alpha} \ga^j \ot 
(\h{\pi}(\omega^j_{\alpha}) + \tJ{j}{n})) 
:= \dsum_{\alpha} \dsum_{j=0}^n (f c^{n-j}_{\alpha}) 
\ga^j \ot (\h{\pi}(\omega^j_{\alpha}) + \tJ{j}{n})~,  \raisetag{1.5ex}
	}
for $f \in \CX\,.$ Moreover, the Hilbert space $h=L^2(X,S) \ot \C^F$ carries 
a natural $\Gamma^\infty(C)$--module structure induced by the 
$\Gamma^\infty(C)$--module structure of $L^2(X,S)\,:$
\eq{
s^c(\tsum_{\alpha} s_{\alpha} \ot \varphi_{\alpha}) :=
\tsum_{\alpha} s^c s_{\alpha} \ot \varphi_{\alpha}~, \quad  
s^c \in \Gamma^\infty(C)~,~~ s \in L^2(X,S)~,~~\varphi \in \C^F~.~~
   }
The structures just introduced enable us to restrict the set of connections 
according to Definition~\ref{connect} to the subset of local connections 
relevant for physical applications.
\begin{dfn}
\label{loccon}
A connection $(\nabla,\nabla_h)$ is called local connection iff for all 
$f \in \CX\,,$ $\bsj \in h$ and $\vrr^n \in \D{n}$ different from boundary 
spaces one has 
\seq{
\eqa{rcl}{
\nabla_h (f \bsj) &=& f \nabla_h(\bsj) + \bfd f (\bsj) ~,~~ \yn \npb \\
\nabla (f \vrr^n) &=& f \nabla(\vrr^n) + (\bfd f) \h{\wedge} \vrr^n~.~~  \yn 
\label{locc}
	 }
The group of local gauge transformations is the group
\eqas[lgg]{rl}{
\mathcal{U}_0(\f[g]):=\big\{\; & u \in \mathcal{U}(\f[g]) \subset \B(h)\;,~
f u \bsj = u f \bsj \;,~ \forall f \in \CX\;,~ \forall \bsj \in h ~,~~ \\
& (\Ad{u} \nabla \Ad{u^*}\,,\, u \nabla_h u^*)~ \mbox{ is a local connection if 
$(\nabla,\nabla_h)$ is}~\big\}\;.~~
	}
        }
\end{dfn}

\subsubsection{Local Connection Forms}

We recall that a connection has the form $(\nabla=d+[\t{\rho},~.~]_g\,,\, 
\nabla_h=-\iu D+\rho)\,,$ where $\rho \in \H{1}$ and $\t{\rho} 
:= \rho + \tcc{1} \in \hH{1}\,,$ see Proposition~\ref{ltheta}. The insertion 
into Definition~\ref{loccon} yields
\eq{
\rho \circ f = f \circ \rho~, \quad \forall f \in \CX~. 
   }
Therefore, $\rho \in \Gamma(C) \ot \mat{F}\,.$ Since $\rho \in \H{1}\,,$ there 
can only occur classical smooth differential forms up to first degree in the 
$\Gamma(C)$--component of $\rho\,.$ This means that
\eqa{rl}{
\t{\rho} \in & (\Lambda^1 \ot \bbr{0}) \op (\Lambda^0 \ga \ot \bbr{1})~,~~ \yn 
\label{rlh} \npb \\ &
\bbr{0} = -(\bbr{0})^* = \h{\Gamma} (\bbr{0}) \h{\Gamma} \subset \mat{F}~, 
\qquad
\bbr{1} = -(\bbr{1})^* = -\h{\Gamma} (\bbr{1}) \h{\Gamma} \subset \mat{F}~.~~
        }
If we compute graded commutators with $\P{*}$ we get 
\al{
[\bbr{0} , \pf] &\subset \pf~, & [\bbr{0} , \p{1}] &\subset \p{1} ~, 
\label{RLH}\\
\{\bbr{0} , \pf\} &\subset \{ \pf, \pf\} + \p{2}~, &
\{\bbr{0} , \p{1}\} &\subset \{ \pf, \p{1}\} + \p{3}~, \notag\\{}
[\bbr{1} ,\pf] &\subset \p{1}~, &
\{\bbr{1} ,\p{1}\} &\subset \p{2} + \{ \pf, \pf\} ~.  \notag
}
Moreover, one has to check that $[\rho,\J{n}]_g \subset \J{n+1}\,.$ Comparing 
this formula with Theorem~\ref{maint}, we must demand
\eqa{rclrcl}{
[\bbr{0} , \jj{k}] &\subset& \tJ{k}{N+k}~, \qquad{} & 
[\bbr{0} , \h{\sigma} \circ \h{\pi}^{-1} ( \K{k}{n} \cap \p{k+2})] 
&\subset& \tJ{k+1}{N+k+1}~, \\{} 
\{ \bbr{0} , \jj{k}\} &\subset& \tJ{k+2}{N+k+2}~, \qquad{} &
\{\bbr{0} , \h{\sigma} \circ \h{\pi}^{-1} ( \K{k}{n} \cap \p{k+2})\} 
&\subset& \tJ{k+3}{N+k+3}~, \\{} 
[\bbr{1} , \jj{k}]_g &\subset& \tJ{k+1}{N+k+1}~, \qquad{} &
[\bbr{1} , \h{\sigma} \circ \h{\pi}^{-1} ( \K{k}{n} \cap \p{k+2})]
&\subset& \tJ{k+2}{N+k+2}~, \npb \\ \yn \label{RJH} 
  }
for all $k,n \in \N\,.$ \hfill The remaining commutators and anticommutators 
$[\bbr{0}, \K{k}{n}]\,,$ $\{\bbr{0}, \K{k}{n}\}$ and $[\bbr{1}, \K{k}{n}]_g$ 
can always be transformed into $[\pf, \K{k}{n}]\,,$ $\{\pf, \K{k}{n}\}$ and 
$[\p{1}, \K{k}{n}]_g$ by means of \rf[RLH]. 

\subsubsection{Local Curvatures}

From \rf[locc] one easily finds for the curvature of a local connection 
$\nabla^2 f = f\nabla^2\,,$ for $f \in \CX\,.$ Thus, 
\eqa{rcl}{
f \theta = \theta f &=& f( \pi \circ d \circ \pi^{-1} (\rho) 
+ \th [\rho,\rho]_g + \J{2} + \tcc{2} ) \npb \\
&=& (\pi \circ d \circ \pi^{-1} (\rho) + \th [\rho,\rho]_g + \J{2} 
+ \tcc{2}) f~.~~ \yn\label{qffq}
   }
Here, $\pi \circ d \circ \pi^{-1} (\rho) + \J{2} + \tcc{2}$ is understood in 
the sense \rf[dhk]. Hence, we must search for the subspace of $\tcc{2}$ 
commuting with functions. This space has the structure
\eq[cja]{
\tcc{2} = (\Lambda^2 \ot \cc{0}) \op (\Lambda^1 \ga \ot \cc{1}) 
\op (\Lambda^0 \ot \cc{2}) {}~,\quad \cc{i} \subset \mat{F}~,~~  
        }
because possible $\Lambda^*$--contributions of higher degree are already 
orthogonal to any representative of $\theta\,,$ see \rf[fee]. The spaces 
$\cc{i}$ have elementwise the following involution and $\Z_2$--grading 
properties:
\eqas[cjb]{rcrcrc}{
\cc{0} &=& -(\cc{0})^* &=& \h{\Gamma} (\cc{0}) \h{\Gamma}\;, \qquad {} &
\cc{1}=-(\cc{1})^*= -\h{\Gamma} (\cc{1}) \h{\Gamma}\;, \\
\cc{2} &=& (\cc{2})^* &=& \h{\Gamma} (\cc{2}) \h{\Gamma}\;. \qquad{}
        }
From \rf[cc] one finds after a decomposition into $\Lambda^*$--components the 
equations 
\seq{
\label{cj0}
\eqas[cjjn]{rclrcl}{
\cc{0} \cdot \pf[a'] &=& 0~, \qquad \qquad{} & \cc{0} \cdot \p{1} &=& 0 ~, \\
\cc{1} \cdot \pf[a'] &=& 0~, \qquad \qquad{} & \cc{1} \cdot \p{1} &=& 0 ~, \\{}
[\cc{2}, \pf[a']] &=& 0~, \qquad\qquad{} & [\cc{2}, \p{1} ] &=& 0~. 
	    }
The restriction to $\pf[a']$ is due to possible problems with the boundary 
spaces. Due to \rf[qffq] it is convenient to define 
\eq{
\cj{0} := \cc{0}~, \qquad  \cj{1} := \cc{1}~, \qquad
\cj{2} := \cc{2} + \jj{2} + \{\pf,\pf\} ~.  \label{cj2cc}
  }
  }

We recall that the commutator and the differential in the curvature 
$\theta=d \t{\rho} + \th [\t{\rho},\t{\rho}]_g$ are indirectly defined via the 
graded Jacobi identity and the graded Leibniz rule \rf[dhk]. The commutator and 
differential in $\P{*} \mod \J{*}$ are given by \rf[vst] and \rf[dt1]. It is 
obvious that these formulae extend to local elements of $\hH{*}\,.$ Only the 
map $\h{\sigma} \circ \h{\pi}^{-1}$ has to be extended to $\bbr{*}$ via the 
graded Leibniz rule:
\eqa{rl}{
[\h{\sigma} \circ \pi^{-1}(\eta^k) & + \jj{k+1}, \h{\pi}(\omega^l) + \jj{l} ]_g 
\npb \\
:= & \h{\sigma} \circ \pi^{-1} ([\eta^k, \h{\pi}(\omega^l)]_g) -(-1)^k 
[\eta^k, \h{\sigma}(\omega^l)]_g + \tJ{k+l+1}{*} ~,~~ \yn \label{hse}
  }
for $\eta^k \in \bbr{k}$ and $\omega^l \in \W[a]{l}\,.$ Then we find for the 
curvature
\eqa{rl}{
\theta = \big( & (\bfd \ot \one_F)(\rho) + \{ \ga \ot -\iu \M,\rho \} 
+ \th \{\rho,\rho\} \npb \\ & 
+ (1 \ot \h{\sigma} \circ \pi^{-1}) \circ \rho \circ (\ga \ot \one_F) \big) 
\mod \cJ{2}~,~~ \label{expq} \yn
   }
where we recall that $\cJ{2} = \Lambda^0 \ot (\jj{2} + \{\pf,\pf\} ) 
+ \tcc{2}\,.$

\subsubsection{The Group of Local Gauge Transformations}

The analysis of the group of local gauge transformations \rf[lgg] yields 
\eqas[ufg]{rcl}{
\mathcal{U}_0(\f[g]) &=& \exp( \Lambda^0 \ot \mathbbm{u}_0(\f[a]) )~, \qquad 
\mbox{where} \\
\mathbbm{u}_0(\f[a]) &=& \{~u_0 \in \bbr{0}~,\quad 
\h{\sigma} \circ \h{\pi}^{-1}(u_0) \subset \cc{1}~\}~,
  }
see \rf[bbmu] and \rf[pex]. 

\subsubsection{Bosonic and Fermionic Actions}

In our case -- $h=L^2(X,S) \ot \C^F$ -- we have 
$\B(h)= \B(L^2(X,S)) \otimes \mat{F}\,.$ Then, the parameter $\mathrm{d}$ in 
\rf[Dix] is equal to the dimension $N$ of the manifold $X$, see \cite{ac}. 
Moreover, the trace theorem of Alain Connes \cite{ac} says that in this case we 
have 
\eq[tth]{
\Tr_{\omega} ((s^c \ot \mathrm{m})\,|D|^{-N}) = \dfrac{1}{(\frac{N}{2})! 
(4\pi)^{\tfrac{N}{2}}}\, \int_X \! \vg \,\tr_c (s^c) \, \tr(\mathrm{m})~,~~ 
	}
where we recall that $\vg$ denotes the canonical volume form on $X\,,$ 
$\tr_c$ denotes the trace in the Clifford algebra 
$\mathrm{Cliff}_{\!\C}\,(\R^N)\,,$ normalized by $\tr_c(1)= 2^{N/2}\,,$ and 
$\tr(\mathrm{m})$ is the matrix--trace of $\mathrm{m} \in \mat{F}\,.$ We use 
the trace theorem \rf[tth] for the construction of $\f[e](\theta)\,,$ see 
\rf[fee]. For the curvature $\theta$ of a local connection we have according 
to the above considerations a decomposition
\eq[qq]{
\theta=\tsum_{\alpha} \big( c^2_{\alpha} \ot (\tau^0_{\alpha} + \cj{0}) + 
c^1_{\alpha} \ga \ot (\tau^1_{\alpha} + \cj{1}) 
+ c^0_{\alpha} \ot (\tau^2_{\alpha} + \cj{2}) \big)~,~~
       }
where $c^i_{\alpha} \in \Lambda^i$ and $\tau^i_{\alpha} \in \mat{F}\,.$ Since 
$\Lambda^*=\bigoplus_{k=0}^N \Lambda^k$ is an orthogonal decomposition with 
respect to the scalar product \rf[ix] given by $\tr_c\,,$ we see that \rf[fee] 
is equivalent to finding for $i \in \{0,1,2\}$ and each $\alpha$ the elements 
$j^i_{\alpha} \in \cj{i}$ satisfying
\seq{
\label{trjj}
\eq[trj]{
\tr( \t{j}^{\,i} \, (\tau^i_{\alpha} + j^i_{\alpha} ))=0~,~~ \mbox{ for all }~
\t{j}^{\,i} \in \cj{i} ~. ~~ 
	}
These equations must be solved depending on the concrete element 
$\tau^i_{\alpha}$ and the concrete L--cycle 
$(\f[a],\C^F, \M,\h{\pi},\h{\Gamma})\,,$ giving in the notation of \rf[qq]
\eq{
\f[e](\theta) = \tsum_{\alpha} \big( c^2_{\alpha} 
\ot (\tau^0_{\alpha} + j^0_{\alpha}) + c^1_{\alpha} \ga \ot (\tau^1_{\alpha} 
+ j^1_{\alpha}) + c^0_{\alpha} \ot (\tau^2_{\alpha} + j^2_{\alpha}) \big)~.~~ 
\yn
    }
    }
Now, formula~\rf[actb] for the bosonic action takes the form (up to a constant)
\seq{
\eq[sbc]{
S_B(\nabla) = \int_X \!\! \vg \; \tr_c ( \f[e](\theta)^2 )~.~~	
	}
Here, $\tr_c$ contains both the traces in $\mathrm{Cliff}_{\!\C}\,(\R^N)$ and 
$\mat{F}\,.$ For the fermionic action we obtain 
\eq[sfv]{
S_F(\bsj,\nabla) = \langle \bsj , (D+\iu \rho) \bsj \rangle_h
= \int_X \!\! \vg \; \bsj^* (D+\iu \rho) \bsj ~.~~ 
	}
        }

\subsubsection{Summary}

This finishes our prescription towards gauge field theories. Let us recall what 
the essential steps are. One starts to select the L--cycle from the physical 
data or assumptions. We have learned that the matrix part of the L--cycle 
contains the essential information. Hence, we must construct the spaces 
$\p{n}$ and the ideal $\jj{n}$ at least up to second and at most up to fourth 
order. This is necessary to compute the spaces $\bbr{0},\bbr{1}$ and 
$\cj{0},\cj{1},\cj{2}$ constituting the connection form $\rho$ and the ideal 
$\cJ{2}\,.$ Then we have to compute the curvature $\theta$ of the connection 
and to select its representative $\f[e](\theta)$ orthogonal to $\cJ{2}\,.$ 
Finally, we write down the bosonic and fermionic actions. This scheme can be 
applied to a large class of physical models. Among them are the 
$\SU3 \times \SU2 \times \U1$--standard model \cite{rw3} and the flipped 
$\SU5 \times \U1$--Grand Unification model \cite{rw4}.


\begin{thebibliography}{99} 
\hbadness1500
\parskip0ex

\bibitem{bgv} N.~Berline, E.~Getzler and M.~Vergne, \emph{Heat Kernels and 
Dirac Operators}, \\ Springer--Verlag Berlin Heidelberg 1992.

\bibitem{cff1} A.~H.~Chamseddine, G.~Felder and J.~Fr\"ohlich, 
\emph{Grand Unification in Non-Com\-mu\-ta\-tive Geometry}, 
Nucl.\ Phys.\ \textbf{B~395} (1992) 672--698.

\bibitem{cff2} A.~H.~Chamseddine, G.~Felder and J.~Fr\"ohlich, \emph{Unified 
Gauge Theories in Non-Commutative Geometry}, Phys.\ Lett.\ \textbf{B~296} 
(1992) 109--116.

\bibitem{cf} A.~H.~Chamseddine and J.~Fr\"ohlich, \emph{$SO(10)$ Unification in 
Non--Commuta\-tive Geometry}, Phys.\ Rev.\ \textbf{D~50} (1994) 2893--2907. 

\bibitem{ac} A.~Connes, \emph{Non commutative geometry}, Academic Press, 
New York 1994. 

\bibitem{acr} A.~Connes, \emph{Noncommutative geometry and reality}, 
J.\ Math.\ Phys.\ \textbf{36} (1995) 6194--6231. 

\bibitem{cl} A.~Connes and J.~Lott, \emph{The Metric Aspect of Noncommutative 
Geometry}, Proceedings of 1991 Carg\`{e}se Summer Conference, ed.\ by 
J.~Fr\"ohlich et al, Plenum, New York 1992.

\bibitem{g} P.~B.~Gilkey, \emph{Invariance Theory, The Heat Equation, And the 
Atiyah--Singer Index Theorem}, Publish or Perish, Wilmington 1984.

\bibitem{h} J.~E.~Humphreys, \emph{Introduction to Lie Algebras and 
Representation Theory}, \\ Springer--Verlag, New York 1972.

\bibitem{iks} B.~Iochum, D.~Kastler and T.~Sch\"ucker, \emph{Fuzzy Mass 
Relations in the Standard Model}, preprint hep-th/9507150.

\bibitem{lmms} F.~Lizzi, G.~Mangano, G.~Miele and G.~Sparano, \emph{Constraints 
on Unified Gauge Theories from Noncommutative Geometry}, 
preprint hep-th/9603095.

\bibitem{kppw} W.~Kalau, N.~A.~Papadopoulos, J.~Plass and J.-M.~Warzecha, 
\emph{Differential Algebras in Non--Commutative Geometry}, 
J.\ Geom.\ Phys.\ \textbf{16} (1995) 149--167. 

\bibitem{mgv} C.~P.~Mart\'{\i}n, J.~M.~Gracia--Bond\'{\i}a and J.~C.~Varilly, 
\emph{The standard model as a noncommutative geometry: the low energy regime}, 
preprint hep-th/9605001.

\bibitem{phd} R.~Wulkenhaar, \emph{Non--associative Geometry and Models of 
Grand Unification}, Ph.D.~thesis, in preparation.

\bibitem{rw1} R.~Wulkenhaar, \emph{A Tour through Non--associative Geometry},\\
preprint hep-th/9607086.

\bibitem{rw3} R.~Wulkenhaar, \emph{The Standard Model within Non--associative 
Geometry}, preprint hep-th/9607096.

\bibitem{rw4} R.~Wulkenhaar, Grand Unification in Non--associative 
Geometry, preprint hep-th/9607237.


\end{thebibliography}
\end{document}